\documentclass{emulateapj}
\usepackage{amsmath,amssymb}
\usepackage{float}
\usepackage{booktabs}
\usepackage{color}
\usepackage{natbib}
\usepackage{graphicx}
\newcommand{\ergs}{ergs s\ensuremath{^{-1}}}

\def\gtrsim{\mathrel{\hbox{\rlap{\hbox{\lower4pt\hbox{$\sim$}}}\hbox{\raise2pt\hbox{$>$}}}}}

\newcommand{\kms}{km~s\ensuremath{^{-1}}}

\newcommand{\ledd}{\ensuremath{L\mathrm{_{Edd}}}}

\newcommand{\lbol}{\ensuremath{L_{\mathrm{bol}}}}
\newcommand{\loiii}{\ensuremath{L_{\mathrm{[O {\tiny III}]}}}}

\newcommand{\msun}{\ensuremath{M_{\odot}}}
\newcommand{\lsun}{\ensuremath{L_{\odot}}}

\newcommand{\oiii}{[\ion{O}{3}]}

\newcommand{\degree}{^\circ}

\newcommand{\XCO}{X$_{\rm{CO}}$}
\newcommand{\alphaCO}{$\alpha_{\rm{CO}}$}

\def\lax{{$\mathrel{\hbox{\rlap{\hbox{\lower4pt\hbox{$\sim$}}}\hbox{$<$}}}$}}
\def\gax{{$\mathrel{\hbox{\rlap{\hbox{\lower4pt\hbox{$\sim$}}}\hbox{$>$}}}$}}

\slugcomment{Draft version June 16, 2014}
\shorttitle{Candidate Molecular Outflow in an Obscured Quasar}
\shortauthors{}

\begin{document}

\title{ALMA observations of a candidate molecular outflow in an obscured quasar}

\author{Ai-Lei Sun\altaffilmark{1}\altaffilmark{*}, Jenny E. Greene\altaffilmark{1}, Nadia L. 
Zakamska\altaffilmark{2}, Nicole P. H. Nesvadba\altaffilmark{3}}
\altaffiltext{1}{Department of Astrophysics, Princeton University, Princeton, NJ 08540, USA}
\altaffiltext{2}{Department of Physics and Astronomy, Bloomberg Center, Johns Hopkins University, Baltimore, MD 21218, USA}
\altaffiltext{3}{Institut d'Astrophysique Spatiale, CNRS, Universit\'{e} Paris-Sud, Bat. 120-121, 91405 Orsay, France}
\altaffiltext{*}{E-mail: aisun@astro.princeton.edu}

\begin{abstract}

We present Atacama Large Millimeter/Submillimeter Array (ALMA) CO (1-0) and CO (3-2) observations of SDSS J135646.10+102609.0, an obscured quasar and ultra-luminous infrared galaxy (ULIRG) with two merging nuclei and a known 20-kpc-scale ionized outflow. The total molecular gas mass is $M_{\rm{mol}}\approx9^{+19}_{-6}\times 10^8$ \msun, mostly distributed in a compact rotating disk at the primary nucleus ($M_{\rm{mol}}\approx3\times 10^8$ \msun) and an extended tidal arm ($M_{\rm{mol}}\approx5\times 10^8$ \msun). The tidal arm is one of the most massive molecular tidal features known; we suggest that it is due to the lower chance of shock dissociation in this elliptical/disk galaxy merger. 
In the spatially resolved CO (3-2) data, we find a compact ($r \approx 0.3$ kpc) high velocity ($v\approx500$ \kms) red-shifted feature in addition to the rotation at the N nucleus. 
We propose a molecular outflow as the most likely explanation for the high velocity gas. 
The outflowing mass of $M_{\rm{mol}}\approx7\times 10^7$ \msun~and the short dynamical time of $t_{\rm{dyn}}\approx 0.6 $ Myr yield a very high outflow rate of $\dot{M}_{\rm{mol}}\approx 350$ \msun~yr$^{-1}$ and can deplete the gas in a million years. 
We find a low star formation rate ($< 16$ \msun~yr$^{-1}$ from the molecular content and $< 21$ \msun~yr$^{-1}$ from the far-infrared spectral energy distribution decomposition) that is inadequate to supply the kinetic luminosity of the outflow ($\dot{E}\approx3\times10^{43}$ \ergs).  Therefore, the active galactic nucleus, with a bolometric luminosity of $10^{46}$ \ergs, likely powers the outflow.
The momentum boost rate of the outflow ($\dot{p}/(\lbol/c)\approx$3) is lower than typical molecular outflows associated with AGN, which may be related to its compactness. 
The molecular and ionized outflows are likely two distinct bursts induced by episodic AGN activity that varies on a time scale of $10^7$ yr.

\end{abstract}

\section{Introduction}

In the past decade, supermassive black holes (BHs) have been found to be
a common constituent of galaxy centers \citep[e.g., ][]{2013ARA&A..51..511K}. At the same time, we have come
to appreciate that supermassive BHs may play an active role in shaping galaxy
evolution through feedback in the active galactic nucleus (AGN) phase \citep[e.g.,][]{2012ARA&A..50..455F}.
The steep high-mass cut-off in the galaxy luminosity function \citep[e.g., ][]{2006MNRAS.370..645B} and the 
over-production of massive blue galaxies in cosmological simulations \citep[e.g., ][]{2006MNRAS.365...11C}
requires a mechanism to quench star formation at the high-mass end. The
enormous amount of energy released by black hole accretion provides a way to heat up or expel gas on galaxy-wide scales \citep[e.g., ][]{2001ApJ...551..131C, FaucherGiguere:2012eq}. This kind of AGN feedback may regulate star formation \citep[e.g., ][]{2012Natur.485..213P} and/or BH growth %\citep{1998A&A...331L...1S, 2005Natur.433..604D,
\citep[e.g., ][]{2010MNRAS.405L...1B} in a way that links the evolution of the
BH and the galaxy and results in the local BH scaling relations
\citep[e.g., ][]{2013ApJ...764..184M,2013ARA&A..51..511K,2013ApJ...778...47S}. 

The actual mechanisms that link the AGN with galaxy-scale gas remain unknown. 
There have been many studies searching for  outflows in different gas phases, including X-ray emitting hot gas 
\citep[e.g.,][]{2009ApJ...704.1195W,2011ApJ...736...62W,2014ApJ...788...54G}, warm ionized \citep[e.g.,][]{1987ApJ...316..584S, 1992ApJ...387..109W, Nesvadba:2006jt, 2009ApJ...690..953F, 2011ApJ...732....9G, 2011MNRAS.418.2032V, 2013ApJ...768...75R, 2013ApJ...779...53Y,2013MNRAS.430.2327L, Hainline:2013ek, 
2014MNRAS.440.3202V, _2014arXiv1402.6736Z}, neutral \citep[e.g.,][]{2005ApJ...632..751R,2013ApJ...765...95T}, and cold molecular gas \citep[e.g.,][]{2002ApJ...580L..21W,2010A&A...518L.155F,2011ApJ...735...88A,2011ApJ...733L..16S,Aalto:2012em,Flower:2013gm,2013A&A...549A..51F, 2013ApJ...776...27V,2014A&A...562A..21C}. 
As has been shown by these studies, AGN-driven outflow is a complex phenomenon involving gas at a wide range of temperatures, densities, and distributions. Identifying the driving mechanism of the outflow, e.g. star formation or AGN, is challenging in many cases, in particular because of ambiguities between star formation and AGN activity indicators. 
Therefore, multi-wavelength observations are essential to put together a comprehensive picture of AGN feedback. 

Only recently have we come to appreciate the ubiquity of molecular outflows. They are commonly seen in \emph{Herschel} OH spectroscopy \citep[e.g.,][]{2011ApJ...733L..16S, 2013ApJ...776...27V}. 
Also, exciting evidence from interferometric observations suggest that these molecular outflows are massive, with a high mass loss rate that can deplete the cold gas in the galaxy in $10^6-10^8$ yr \citep[e.g.,][]{2010A&A...518L.155F, 2011ApJ...735...88A,2014A&A...562A..21C}. 
As molecular gas is the fuel for star formation, monitoring the molecular content in the host galaxy is 
key to determining whether or not the AGN can quench star formation and shape galaxy evolution.  

In this paper, we inspect the molecular properties of the obscured
quasar \\ SDSS J135646.10+102609.0 (SDSS J1356+1026 hereafter) with the
Atacama Large Millimeter/Submillimeter Array  (ALMA),
looking for traces of the impact of the AGN on the cold gas component in the
galaxy. SDSS J1356+1026 is an example of feedback in action; an extended 
and energetic outflow is detected in ionized gas that is most 
likely AGN-driven \citep{2012ApJ...746...86G}. 
The spatially resolved ALMA CO (1-0) and CO (3-2) observations presented here allow us to constrain the morphology, kinematics, and mass of the molecular gas and to investigate the relation between the molecular gas and the ionized outflow. 

Throughout, we assume $h=H_0/100
$~\kms ~Mpc$^{-1}$ = 0.7, $\Omega_{\rm{m}} = 0.3$, and
$\Omega_{\mathrm{\Lambda}} = 0.7$.  
At the redshift of the object
(z=0.1231), 1\arcsec~ corresponds to 2.2 kpc, and the luminosity
distance is 580 Mpc. All velocities used in this paper are in the
heliocentric frame using the optical velocity convention. Wavelengths are
expressed in vacuum.

\begin{figure*}[]
        %\centering
        \hbox{
	\includegraphics[bb = 20 110 500 800,clip, scale=0.5]{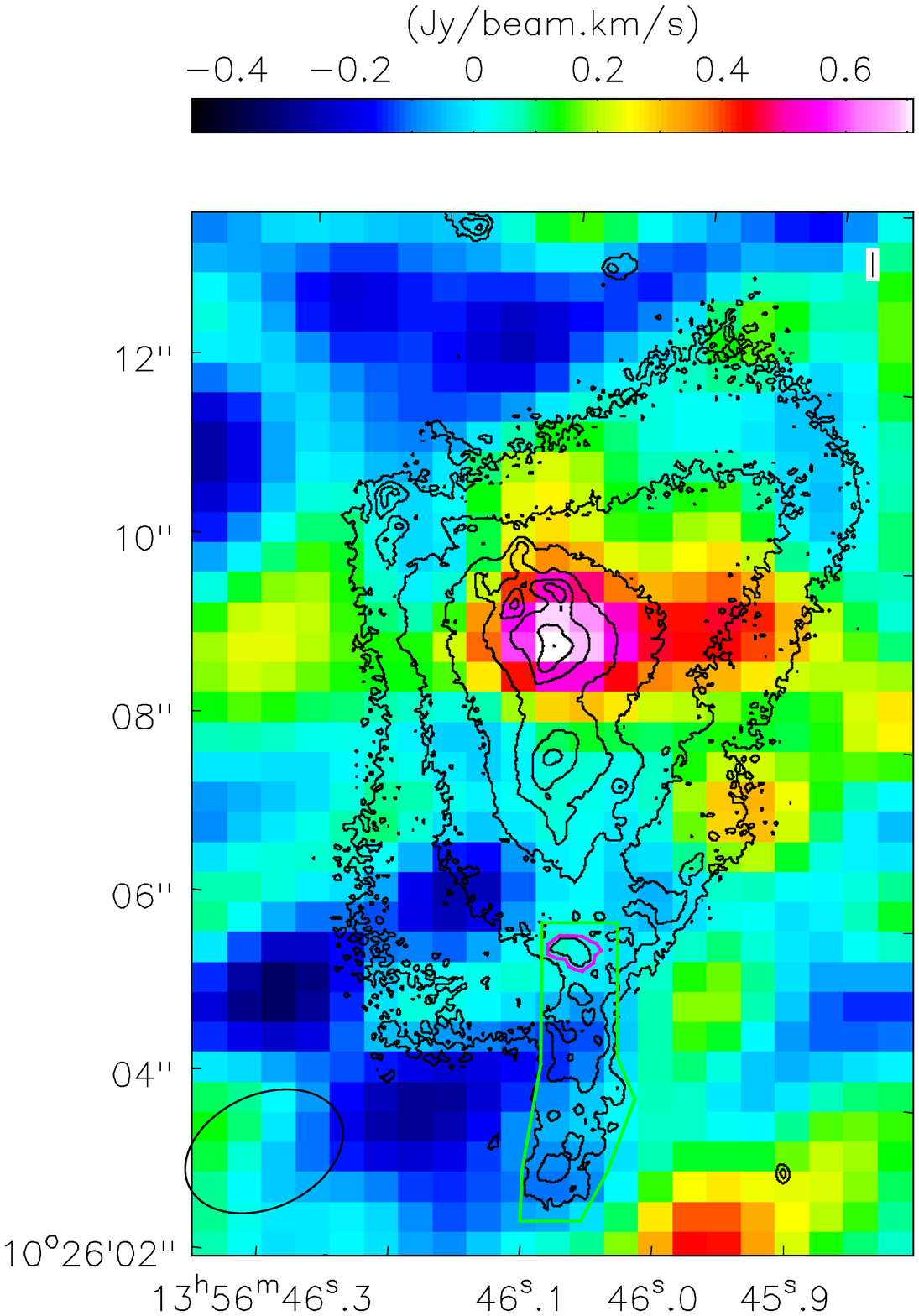}
	\includegraphics[bb = 20 110 500 800,clip, scale=0.5]{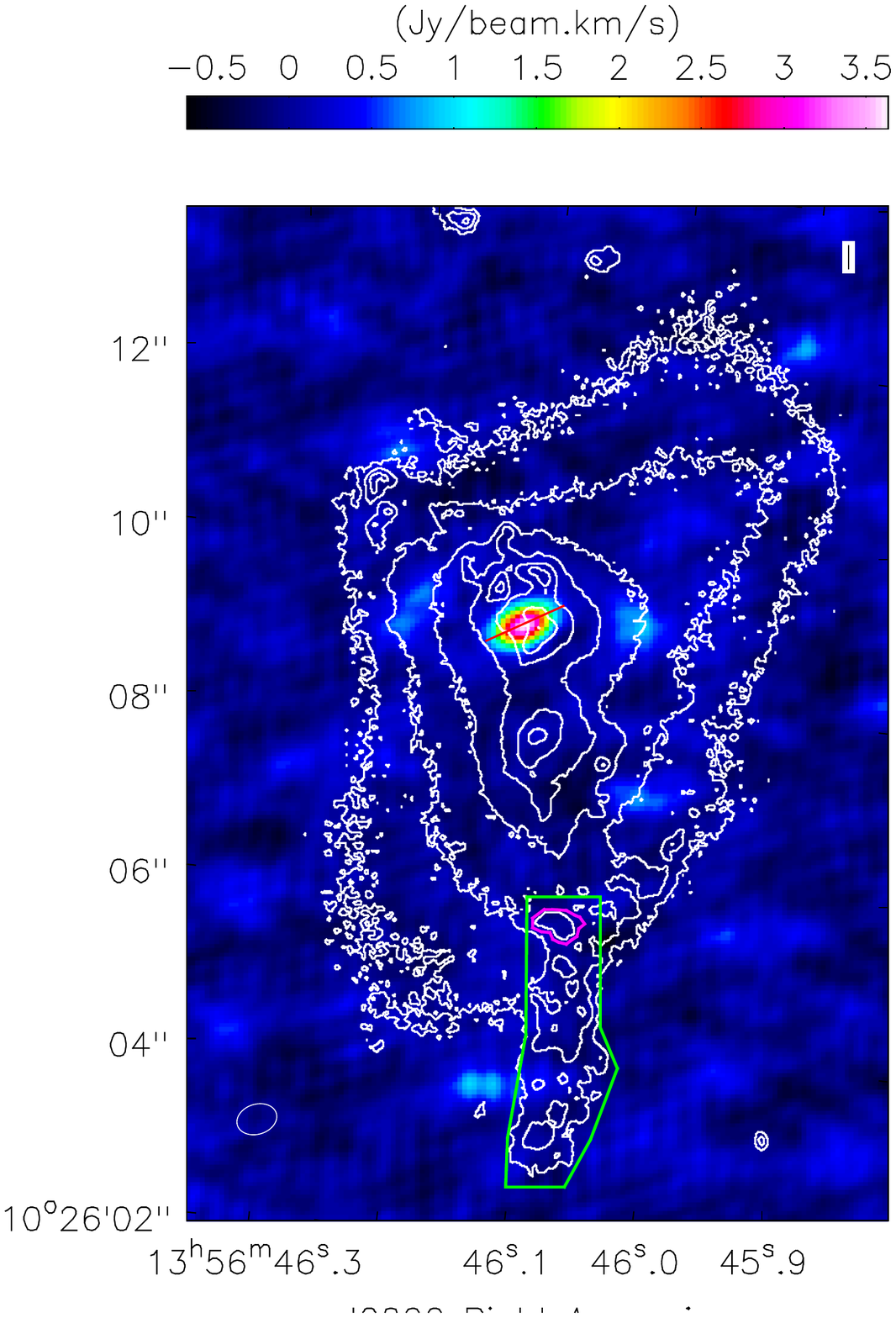}
	}
        \caption{
        Integrated line intensity (moment-0) maps for SDSS J1356+1026 in CO (1-0) (left) and CO (3-2) (right) are shown in color.  These moment-0 maps are integrated over a velocity range of $-300$ to $+300$ \kms. $1~\sigma$ noise level is $0.15$ Jy beam$^{-1}$ \kms~and 0.16 Jy beam$^{-1}$ \kms~for CO (1-0) and CO (3-2) respectively.  The beam ellipse is shown in the lower left corner. Contours show intensity of optical emission as seen in the \emph{HST}/WFC3 F814W $I-$band images and are spaced by a factor of two in intensity.  
        The bold letters in the left panel indicate the three major components, the N/S nucleus and the W arm. The CO emission is most prominent at the N nucleus and the W arm $2\arcsec$ to the west of the N nucleus. Emission at the S nucleus is also detected in CO (1-0) in certain channels, see Fig. \ref{fig:spec}. The red clump at the bottom of the CO (1-0) map is noise at 3 $\sigma$ level.
        The green region marks the the ionized outflow, and the magenta region is its base. There is no detection of a CO counterpart to the extended (10 kpc) ionized \oiii~outflow. The red line across the N nucleus on the right panel indicates the $pv$-diagram cut (Fig. \ref{fig:pv}). 
        }
        \label{fig:moment0}
\end{figure*}

\subsection{SDSS J1356+1026}
\label{sec:intro_bubble}

SDSS J1356+1026 is a merging system at z=0.123 classified as a
luminous obscured (Type 2) quasar with a bolometric luminosity \lbol
$\approx 10^{46}$ \ergs~% = $3\times10^{12}$ \lsun~ %($\pm$0.5 dex) 
inferred from the \oiii~
luminosity, \loiii = $10^{42.77}$~\ergs~% = $1.5\times10^9$ \lsun~
\citep{Greene:2009kt,Liu:2009gr}. It is 
also an ultra-luminous infrared galaxy (ULIRG; $L_{\rm{FIR}} =2.68 \pm 0.53 \times10^{45}$ \ergs). The radio continuum is unresolved with a luminosity density of $\nu L_{\nu}=3.4\times10^{40}$ \ergs~from the VLA FIRST survey at 1.4 GHz (1.6 GHz rest frame), indicating that it is a radio-quiet quasar \citep{2012ApJ...746...86G}. The coordinate of SDSS J1356+1026 is (13:56:46.10, +10:26:09.09).

The two merging galaxies, which we refer to as the northern (N)
and southern (S) nuclei, are separated by 2.5 kpc ($1\farcs1$). 
The N nucleus is detected at 2-10 keV and is the primary AGN host in the system (Greene et al. 2014). Both the total molecular mass (Section
\ref{sec:gas_mass}) and the {\it r}-band light \citep{Greene:2009kt}
are dominated by the northern nucleus, with a ratio of $\sim 4:1$
(N:S).  The progenitor of the northern galaxy is likely a moderately massive
early-type galaxy, with a stellar spectrum that is dominated by an old
stellar population and a stellar velocity dispersion $\sigma_{*} =
206\pm36$ \kms ~\citep{Greene:2009kt}, corresponding to a stellar mass
$M_* \approx 10^{11}$ \msun~ \citep{Hyde:2009ig}, and the black hole
mass is estimated to be $M_{\rm BH} \approx 3 \times 10^8$ \msun~ \citep[$\pm 0.38$ dex,][]{2013ApJ...764..184M}. The corresponding Eddington luminosity is
\ledd$= 3.1\times10^{46}$ \ergs~($\pm 0.38$ dex), which yields an Eddington ratio range of 0.1-1, energetic enough to drive a hot wind
\citep[e.g.,][]{2005ARA&A..43..769V}.

The most spectacular feature of this system is an \oiii-emitting
bubble extending $\sim 10$~kpc to the south of the nuclei (green region in Fig. \ref{fig:moment0}), with a
weaker symmetric counterpart to the north. Long-slit spectroscopy
along the ouflow reveals the distinctive double-peaked spectrum
of an expanding shell. A simple geometric model yields a deprojected velocity $v \approx 1000$ \kms, a
dynamical time $t_{\rm{dyn}} \approx 10^7$ yr, and a kinetic luminosity $\dot{E} \approx 10^{44-45}$~\ergs~\citep{2012ApJ...746...86G}. As the radio emission is compact
and faint (Sec. \ref{sec:100GHz}, \ref{sec:driving_source}), and the star formation rate is low (Sec. \ref{sec:gas_mass}, \ref{sec:SED}),
this ionized outflow provides a strong case for a quasar-driven wind \citep{2014ApJ...788...54G}.

\section{Observations}
\subsection{ALMA CO (1-0) and CO (3-2) Observations}
The ALMA CO (1-0) and CO (3-2) observations were conducted during
Cycle 0 and Cycle 1 under project codes 2011.0.00652.S and 2012.1.00797.S 
respectively.  The
CO (1-0), at sky frequency 102.6 GHz, was observed in band 3 in two
blocks on May 9 and July 30, 2012, using 16 and 23 12-m
antennae for 68 and 63 minutes respectively (27 and 24 minutes on-source). The
CO (3-2) at 307.9 GHz was observed in band 7 on July 6, 2013 with
twenty-seven 12-m antennae in 24 minutes (19 minutes on-source). 
In the following, CO (1-0) properties are followed by CO (3-2) in parenthesis. 
Both lines use a single pointing covering the whole system with a field of view
of 62\arcsec~ (21\arcsec), and dual polarizations. They use the same spectral
set-up with a total bandwidth of 2 GHz divided into 128 channels, each
15.625 MHz wide, corresponding to a channel width of 46 \kms~ (15
\kms). The CO (1-0) data has an additional spectral window at 93 GHz. The channels are Hanning-smoothed within the correlator.
 QSO 3C279 (J1516+0015),
Mars (Titan), and J1415+133 (J1347+1217) were used for the bandpass,
flux, and phase calibrators respectively. The phase calibrator is observed for 30
(90) seconds every 11 (9) minutes. 
The absolute flux calibration accuracy is 5\% for CO (1-0) (Band 3) and 10\% for CO (3-2) (Band 7). 
% for both cycle-0 and cycle-1
%

The data calibration was carried out by the ALMA team using CASA
version 3.3.0 for CO (1-0) and 4.1.0 for CO (3-2). We further used
CASA version 4.1.0 to apply continuum subtraction [CO (1-0) only],
heliocentric correction, imaging, and cleaning. For CO (1-0), the
continuum level was estimated from line free channels (36/43 channels
on the blue/red side) and subtracted in the {\it uv}-plane. Continuum
subtraction was not applied to the CO (3-2) data cube as there are not enough
line-free channels to determine the continuum level. 
However, we extrapolate from the 100 GHz and 1.4 GHz continuum flux densities (spectral index $\alpha=-0.86$, Sec. \ref{sec:SED}) to estimate a 300 GHz flux density of 0.58 mJy. This extrapolated continuum level is lower than the noise level in each CO (3-2) channel and is less than 10\% of the emission line flux density at the N nucleus. 
 The CO (3-2) channel is further binned by two to
increase the SNR in each channel, giving a final channel width of 30
\kms, while the CO (1-0) channel width is unchanged (46 \kms). 

The imaging processing was carried out with the CASA task {\it clean}. 
We used the {\it Briggs} visibility weighting with a {\it robustness} parameter of 0.5 \citep{1995AAS...18711202B}, which is 
a compromise between maximum resolution and maximum sensitivity. 
The beam size is $\theta_{\rm{beam}}=1\farcs9 \times1\farcs3~(0\farcs35 \times 0\farcs29)$ with a PA=$-62\degree~(-60\degree)$, 
and the maximum resolvable scale is
$\theta_{\rm{MRS}}= 29\arcsec~(10\arcsec)$.
All images are cleaned to the 3 $\sigma$ level in each
channel. The resulting RMS noise level in each channel is 0.37 mJy beam$^{-1}$ (0.77 mJy beam $^{-1}$).

\begin{deluxetable*}{crcccccc}
\tablecolumns{8} 
\tabletypesize{\scriptsize}
\tablewidth{0pc}
\tablecaption{ Flux Measurements \label{tab:CO10flux}}
\tablehead{
& &\multicolumn{3}{c}{CO (1-0)} &\multicolumn{3}{c}{CO (3-2)}\\
\cmidrule(r){3-5} \cmidrule(r){6-8}
\colhead{Name} & \colhead{$\Delta v$} & \colhead{$S_\nu \Delta v$} & \colhead{$L$} & \colhead{$L'$} & \colhead{$S_\nu \Delta v$} & \colhead{$L$} & \colhead{$L'$}\\
\colhead{} & \colhead{(\kms)} & \colhead{(Jy \kms)} & \colhead{($10^3 \lsun$)} & \colhead{($10^9$ K \kms pc$^2$)} & \colhead{(Jy \kms)} & \colhead{($10^3 \lsun$)} & \colhead{($10^9$ K \kms pc$^2$)}\\
\colhead{(1)} & \colhead{(2)} & \colhead{(3)} & \colhead{(4)} & \colhead{(5)} & \colhead{(6)} & \colhead{(7)} & \colhead{(8)}}
\startdata
N Nucleus & $-300$ : 500 & $0.46\pm0.27$ & $16.6\pm9.6$ & $0.34\pm0.19$&$4.90\pm0.22$&$528\pm24$&$0.40\pm0.02$\\
N Outflow& 300 : 500 & $0.06\pm0.08$ & $2.2\pm2.7$ & $0.04\pm0.06$ &$0.99\pm0.10$&$107\pm11$&$0.08\pm0.01$\\
S Nucleus & $-80$ : 50 & $0.13\pm0.06$ & $4.5\pm2.2$ & $0.09\pm0.04$&$0.11\pm0.07$&$12\pm7$&$0.01\pm0.01$\\
W Arm & $-300$ : 500 & $0.82\pm0.34$ & $29.3\pm12.2$ & $0.60\pm0.25$&$2.55\pm0.40$&$275\pm44$&$0.21\pm0.03$\\
Sum & & $1.40\pm0.44$ & $50.4\pm15.7$ & $1.03\pm0.32$&$7.57\pm0.46$&$815\pm50$&$0.62\pm0.04$
\enddata
\tablecomments{The CO (1-0) and CO (3-2) fluxes and derived properties of the main components (Sec. \ref{sec:flux_measurements}). Columns 3 to 5 pertain to CO (1-0) and columns 6 to 8 pertain to CO (3-2). The last row is a sum of the three components: the N/S nuclei and the W arm. The listed errors correspond to the RMS noise in the data. In addition, there is a 5\% and 10\% flux calibration error for CO (1-0) and CO (3-2) respectively. The CO (1-0) flux errors of the N nucleus and W arm also include errors due to source decomposition. Column 1: component. 
Column 2: the velocity range over which the flux is integrated. Column 3: integrated CO (1-0) flux. Column 4: the CO (1-0) line luminosity, adopting a luminosity distance of 580 Mpc. Column 5: the emitting area and velocity integrated CO (1-0) source brightness temperature. 
Column 6: integrated CO (3-2) flux.
Column 7: the CO (3-2) line luminosity, adopting a luminosity distance of 580 Mpc.
Column 8: the emitting area and velocity integrated CO (3-2) source brightness temperature. 
}
\end{deluxetable*}

\subsection{Matching ALMA with Optical Data}

In order to compare the molecular and stellar components of the
galaxies, we match the ALMA data cube to the SDSS spectrum in velocity
and the Hubble Space Telescope \emph{HST}/WFC3 F814W ($I-$band) image in position (Comerford et al. in prep.).  We use the SDSS DR7 spectrum \citep{2009ApJS..182..543A},
which was taken in a 3$\arcsec$ aperture with a spectral resolution of FWHM $\approx$ 150 \kms~centered on the N nucleus. 
The \emph{HST}/WFC3 F814W image was observed on May
19 2012 with an integration time of 900 seconds and resolution of
$0\farcs07$.
% http://www.stsci.edu/hst/wfc3/documents/handbooks/currentIHB/c06_uvis07.html#391844

% position alignment
To perform positional alignment, we identify another galaxy SDSS
J135646.50+102553.6 that appears in the ALMA, SDSS, and \emph{HST}
images, and two other fainter galaxies in both SDSS and \emph{HST}.  We
found that while the ALMA and SDSS positions are well-matched, there is a
$0\farcs6$ offset from the \emph{HST} position. After applying this offset
to the \emph{HST} image, the ALMA (continuum) and \emph{HST} coordinates of the
northern nucleus are aligned well within $0\farcs04$, much smaller
than the ALMA CO (3-2) beam size ($0\farcs35 \times 0\farcs29$).  Rotation and stretching of the \emph{HST} image with respect to the SDSS
coordinates are also constrained to be within $0.4\degree$ rotation
and $0.5 \%$ stretching.

%velocity alignment
The velocities in this paper are with respect to the rest frame of the stellar absorption features of N galaxy. 
We define the systemic velocity to be the best-fit
redshift (z = 0.1231) from the SDSS spectrum, which is dominated by
the light of the N nucleus, as the N nucleus is brighter and the fiber is centered on it.
 % get the relative intensity of the two nuclei
Both the stellar absorption and AGN emission lines are fitted by the
SDSS template.  A redshift warning flag "many outliers" is raised,
indicating that the template cannot capture the multi-component 
emission lines,
but this is not a major concern for redshift accuracy. As a sanity check,
we refit the stellar continuum using \citet{Bruzual:2003ck} stellar
population synthesis templates and the best-fit redshift differs from
that of the SDSS by $-30$ \kms. We therefore adopt a redshift 
error of $\delta z = \pm 0.0001$ ($\pm$ 30 \kms). This stellar continuum fitting also reveals that the light in the N nucleus is dominated by old stellar populations, as discussed in \citet{Greene:2009kt}.

\begin{figure*}[]
	\centering
	\includegraphics[bb = 70 140 540 690,clip, scale=0.9]{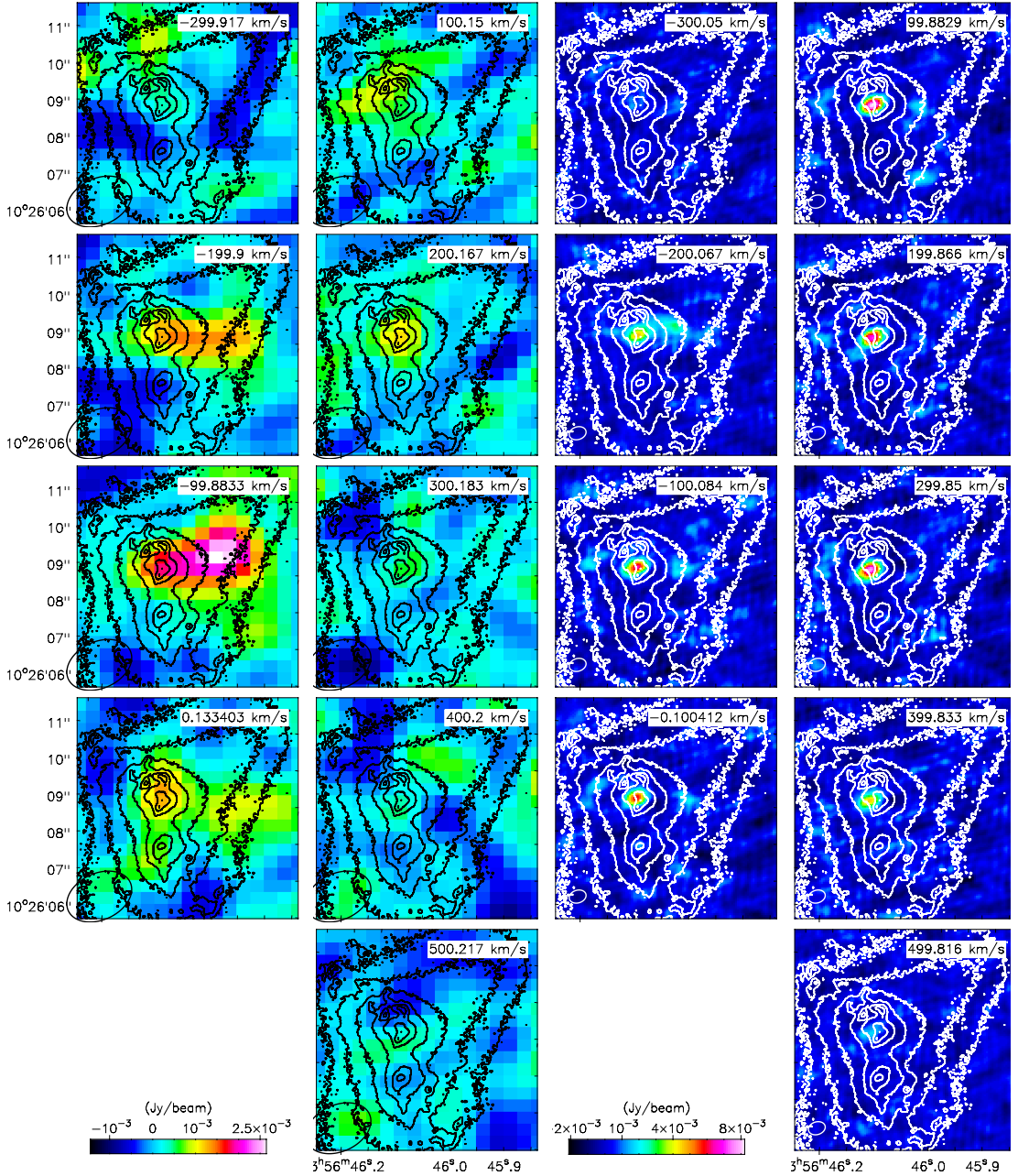}
	 \caption{	 
	 SDSS J1356+1026 CO (1-0) (left) and CO (3-2) (right) channel maps in color. Each of the channel maps is 100 \kms~wide in velocity. The 1-$\sigma$ noise level of the image is 0.31 mJy beam$^{-1}$ \kms~and 0.32 mJy beam$^{-1}$ \kms~for CO (1-0) and CO (3-2) respectively. The beam ellipse is shown in the lower left corner.
	 Contours show intensity of optical emission as seen in the \emph{HST}/WFC3 F814W $I-$band images and are spaced by a factor of two in intensity.   The N nucleus is prominent with a wide velocity range between $-300$ to 500 \kms~especially in the CO (3-2) maps. The W arm is strong in two channels $-200$ and $-100$ \kms, and the S nucleus can be seen at 0 \kms~in the CO (1-0) maps.   }
        \label{fig:channel}
\end{figure*}

\begin{figure*}[]
        \centering
	\includegraphics[scale=0.6]{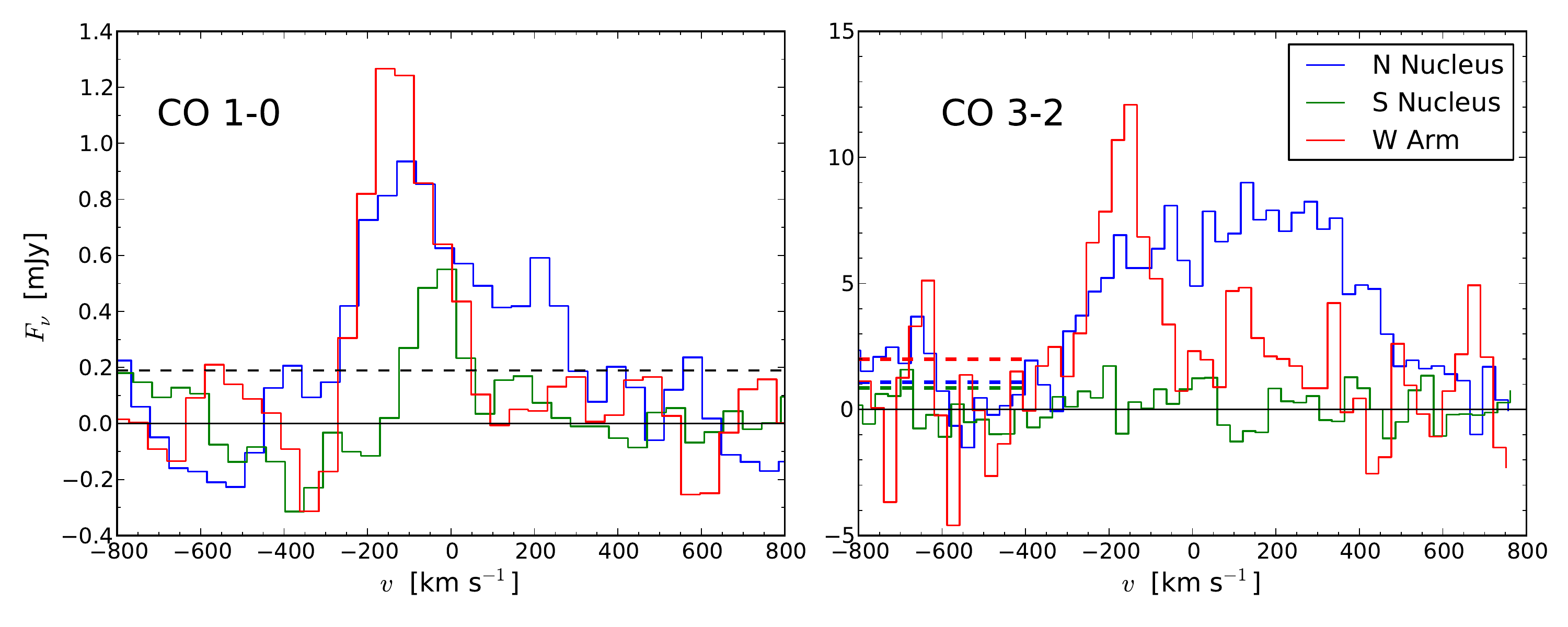}
        \caption{CO (1-0) (left) and CO (3-2) (right) spectra of the N/S nucleus and the W arm. The N nucleus has wide line width in both CO (1-0) and CO (3-2), and the peak at $-100$ \kms~in the CO (1-0) spectrum is contamination from the W arm. The W arm has relatively narrow line width, reflecting its coherent kinematics structure across the arm. The S nucleus is fainter and is detected only in CO (1-0). The dashed horizontal lines indicate the 1 $\sigma$~noise level. To avoid contamination between sources, the CO (1-0) spectra are integrated only over one beam size, and therefore do not include all the flux in the sources. The CO (3-2) spectra are integrated over regions at least twice as large as the beam, and therefore contains $\sim 95 \%$ of the flux in the N nucleus. 
        }
        \label{fig:spec}
\end{figure*}

\begin{figure}[h]
        \centering
	\includegraphics[bb = 47 100 500 780,clip, scale=0.5]{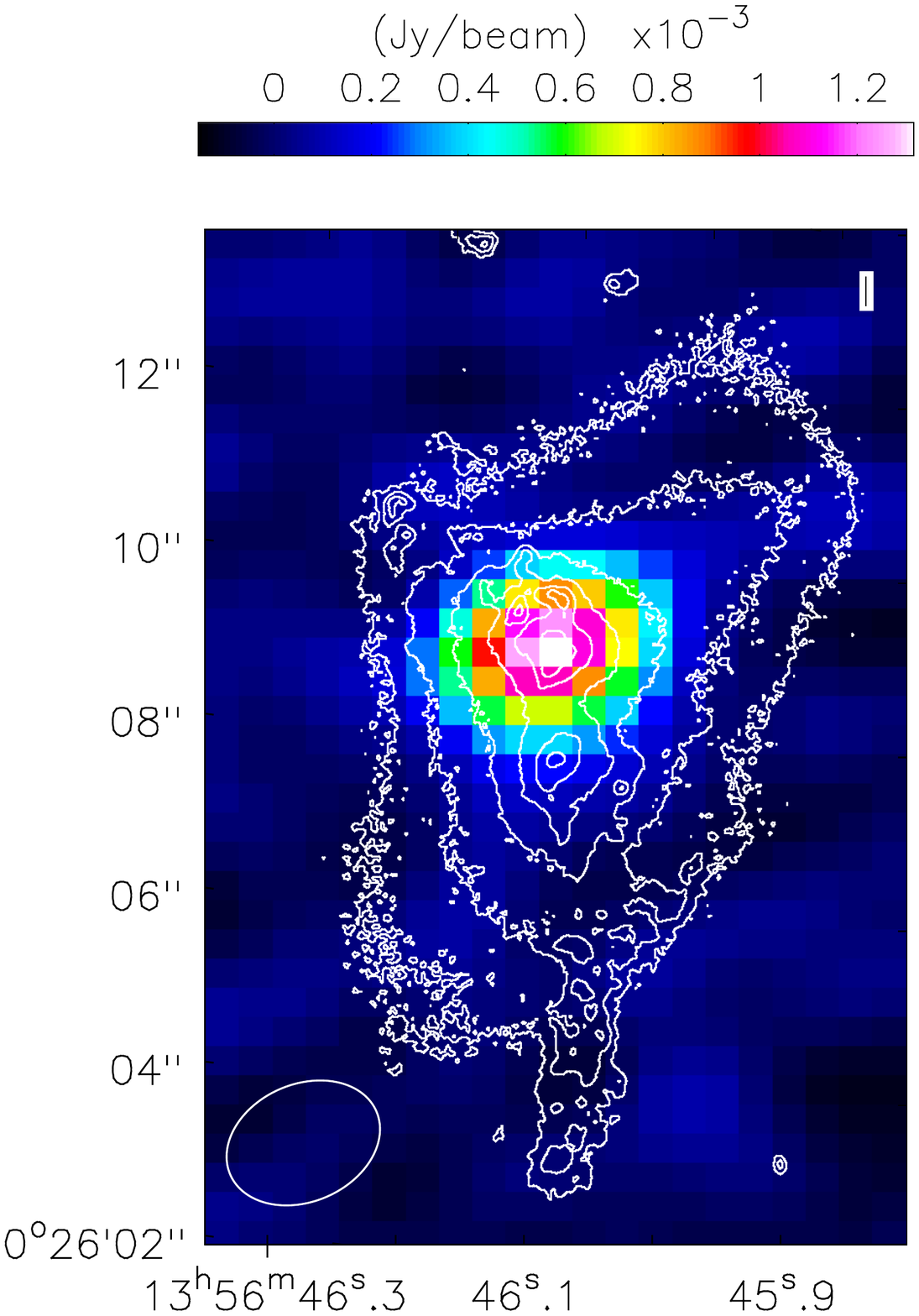}
        \caption{100 GHz radio continuum image (color) overlaid with the \emph{HST}/F814W image (contour). The contours of the \emph{HST} image are in log scale; each contour differs by a factor of 2. The radio image is made from line-free channels at the two spectral windows of 93/103 GHz (rest-frame 104/115 GHz), each 2 GHz wide. There is one unresolved radio continuum source associated with the N nucleus, likely from the AGN (see Fig. \ref{fig:SED}). }
        \label{fig:cont+HST}
\end{figure}

\section{Analysis}
\label{sec:analysis}
% presenting moment - 0, channel maps, and spectrum
In this section, we describe our procedures to estimate the 100 GHz continuum flux density, CO (1-0) and CO (3-2) line fluxes, molecular gas masses, the star formation rate, and the AGN bolometric luminosity. The measurements are summarized in Tables \ref{tab:CO10flux} and \ref{tab:gas_prop}, and will be discussed in Section \ref{sec:results}. 

% introducing components
We first introduce the three major components in the CO data: 
the N nucleus, the S nucleus, and the Western arm (W).  
All three are spatially coincident with optical features.
 The N
nucleus has strong emission from both CO (1-0) and CO (3-2), as seen in the integrated line intensity maps (moment-0, Fig. \ref{fig:moment0}) and the channel maps (Fig. \ref{fig:channel}) with a very broad spectrum (Fig. \ref{fig:spec}). The S nucleus is not detected in CO (3-2), but is seen in CO (1-0) (Fig. \ref{fig:spec}). To the west of the N
nucleus is another strong blue-shifted and extended emission feature in both
transitions, which we call the W arm. This feature overlaps with an
extended stellar plume in the \emph{HST} images, and has blue-shifted but narrow CO lines. Other than these
three main components, no prominent CO emission is detected. There is
no large-scale molecular gas associated with the \oiii\ outflow
discovered by \citet{2012ApJ...746...86G}, but there is a low 
significance (3~$\sigma$)
double-peaked CO (3-2) emission component at the base of the \oiii\
expanding bubble (Sec. \ref{sec:mol_counterpart}). 

\subsection{100 GHz Continuum}
\label{sec:100GHz}

The radio 100 GHz continuum is measured from the line-free channels in two spectral windows at 93/103 GHz (rest-frame 104/115 GHz) in the CO (1-0) data. 
% We use two spectral windows at 93/103 GHz (rest-frame 104/115 GHz) with 116/46 line-free channels. 
The moment-0 map (Fig. \ref{fig:cont+HST}) shows an unresolved point source with a beam size of $1\farcs9 \times1\farcs3$ ($4.2\times2.9$ kpc) at the N nucleus with a flux density of $1.51 \pm 0.07 \pm 0.07$  mJy. The first error is from the RMS noise, and the second is the 5\% calibration error. 
The spectral energy distribution of SDSS J1356+1026 including this 100 GHz flux density is discussed in Section \ref{sec:SED}. 

% 11.2.6 of 
%http://www.iram.fr/IRAMFR/ARC/documents/cycle0/ALMA_TechnicalHandbook_D0.3v1.0.pdf

%with an averaged frequency 97 GHz (rest-frame 109 GHz).  

\subsection{CO Flux Measurements}
\label{sec:flux_measurements}

We extract the CO(1-0)/(3-2) spectra (flux densities $F_{\mathrm{\nu}}$, Fig. \ref{fig:spec}), and fluxes ($F=F_{\mathrm{\nu}}\Delta v$, Table \ref{tab:CO10flux}) for the three components (the N/S nuclei and
the W arm) from the cleaned data.  
In addition to the RMS noise errors presented in Fig. \ref{fig:spec} and Table \ref{tab:CO10flux}, there is a 5 \% and 10 \% flux calibration error for the CO (1-0) and CO (3-2) data respectively. 
We start with the CO (3-2) data as it has higher resolution to spatially separate the components.  
We use simple aperture photometry with
apertures at least twice as large as the beam in order to enclose $\gtrsim 95 \%$ of the flux. As the N and W apertures are adjacent to each other, we expect flux contamination between these two components at a level of 10\% or less for CO (3-2). 
The S aperture is well-separated from the other two sources and does not overlap with side-lobes of the N nucleus.

Assuming Gaussian noise correlated on the scale of the beam, we
estimate the errors in the spectra $F_{\nu}$ based on 
the image noise level rms($I_{\nu}$), taken from large emission-free
regions, 
% and assumed to be a constant across the data cube
the beam $\mathrm{B}(\boldsymbol{r})$, a 2D Gaussian with peak value 1, 
and the aperture $T(\boldsymbol{r}$) = 1 inside the aperture and zero everywhere else, 
adopting the following equation:
\begin{equation}
\mathrm{rms}(F_{\nu})=\mathrm{rms}(I_{\nu})\sqrt{\int\int \mathrm{B}(\boldsymbol{r}_2-\boldsymbol{r}_1)T(\boldsymbol{r}_1)T(\boldsymbol{r}_2) {\rm d}^2\boldsymbol{r}_1 {\rm d}^2\boldsymbol{r}_2 }.
\end{equation}

To obtain the fluxes $F$, we integrate the spectra $F_\nu$ over a velocity range of 
$-300$ \kms $< v<$ 500 \kms\ for the N nucleus and W arm, which is chosen as the velocity width where the CO (3-2) flux density is detected with greater than 2 $\sigma$ significance from the N nucleus. 
A velocity range of 300 \kms $< v<$ 500 \kms\ is used for the high velocity component of the N nucleus. 
For the S nucleus, although there is no CO (3-2) detection, we use the velocity range $-80 < v< 50$ which covers the $> 1~\sigma$ CO (1-0) emission.  While 
estimating the errors in
the fluxes, we take into account the correlation between adjacent channels, which has a Pearson correlation coefficient of $r = 0.375$, due to Hanning smoothing during the observation and 2:1 binning in the image processing.

Estimating the CO (1-0) fluxes 
is more complicated, as the sources, especially the N nucleus and the
W arm, are spatially blended. 
The CO (1-0) spectrum of the N nucleus clearly shows the narrow peak of the W arm at $-100$ \kms~ (Fig. \ref{fig:spec}). However, we can still extract the uncontaminated N nucleus fluxes making use of their distinct spectral shape. 
We first estimate the N nucleus flux in the $100$ to $500$ \kms~channels that are free from W arm emission by fitting a model of the beam to the stacked moment-0 map of these channels.  
This point-source assumption should be valid as the CO (1-0) beam is much larger than the CO (3-2) N nucleus size. 
We then recover the total N nucleus CO (1-0) flux in the entire velocity range ($-300$ to $500$~\kms) assuming that the CO (1-0) spectrum has a similar spectral shape to the CO (3-2) spectrum. 

The W arm CO (1-0) flux is then estimated as the difference between
the total flux including both sources and the decomposed N
nucleus flux. The CO (1-0) flux in the high velocity component of the N nucleus is estimated by the same point source fitting procedure over the velocity range of 300 to 500 \kms, as is the S nucleus over a velocity range of $-80$ to $50$ \kms, where the emission is seen.
Correlations between adjacent channels are determined from emission-free regions and are taken into account for the flux error estimation. 
Using the estimated fluxes of the components in both lines, we express the line ratio as $L'_{\mathrm{CO (3-2)}}/L'_{\mathrm{CO (1-0)}}$ in Table \ref{tab:gas_prop}, where $L'$ is in units of K km s$^{-1}$ pc$^2$, proportional to the brightness temperature.

\begin{figure*}[]
        \centering
        \hbox{
	\includegraphics[scale=0.6]{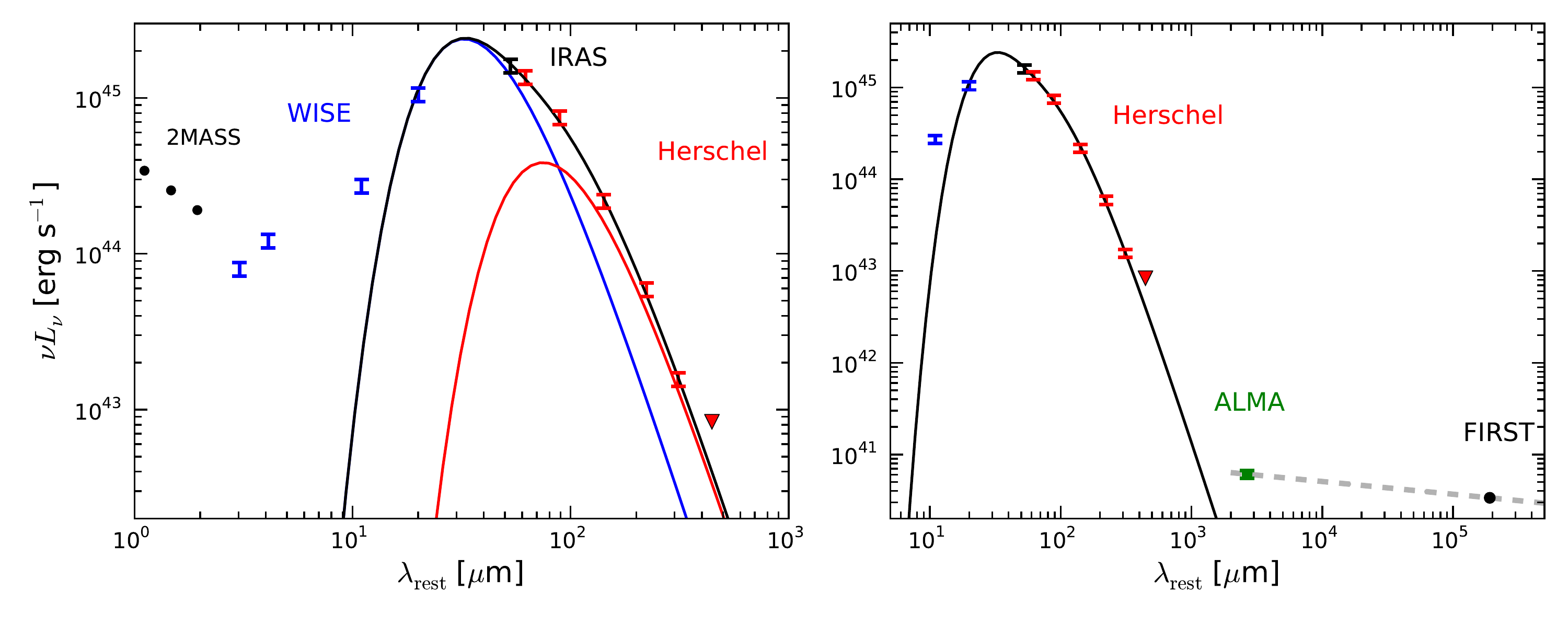}
	}
        \caption{Rest-frame spectral energy distribution of SDSS J1356+1026 (Sec. \ref{sec:SED}, Table \ref{tab:SED}). The black points are taken from the SED summarized by \citet{2012ApJ...746...86G}. In addition, we include the \emph{WISE} (blue), \emph{Herschel} (red), and ALMA 100 GHz (green, Sec. \ref{sec:100GHz}) photometry. The solid black line shows the best-fit two-temperature modified black-body model fitted to the 22 \micron~\emph{WISE}, 60 \micron~\emph{IRAS}, and 70-350 \micron~\emph{Herschel} data. The warm component (blue solid line) has a best-fit temperature of 81 K, while the temperature of the cold component (red solid line) is fixed at 35 K. The grey dashed line on the right panel links the ALMA 100 GHz and FIRST 1.4 GHz \citep[from][]{2012ApJ...746...86G} points with a spectral index of $\alpha=-0.86$~($F_{\nu} \propto \nu^{\alpha}$). }
        \label{fig:SED}
\end{figure*}

\begin{deluxetable}{ccccc}[]
\tablecolumns{5} 
\tabletypesize{\scriptsize}
\tablewidth{0pc}
\tablecaption{Molecular Gas Properties \label{tab:gas_prop}}
\tablehead{
\colhead{Name} & \colhead{$M_{\mathrm{mol}}$} & \colhead{$L'_{\mathrm{CO (3-2)}}/L'_{\mathrm{CO (1-0)}}$}\\
\colhead{} & \colhead{($\msun$)} & \colhead{}\\
\colhead{(1)} & \colhead{(2)} & \colhead{(3)} }
\startdata
N Nucleus&$3^{+6}_{-2}\times 10^8$&$1.18\pm0.73$\\
N Outflow&$7^{+15}_{-5}\times 10^7$$^{~a}$& $>0.45^{~b}$\\ %1.83\pm 2.49~ 
S Nucleus&$<3.8\times10^{8~c}$&$0.10\pm0.11$\\
W Arm&$5^{+11}_{-3}\times 10^8$&$0.35\pm0.20$\\
Sum&$9^{+19}_{-6}\times 10^8$&$0.60\pm0.22$
\enddata
\tablecomments{The molecular gas properties. Column 1: component. Column 2: the molecular gas mass (Sec. \ref{sec:gas_mass}). The errors are dominated by the 0.5 dex error in the \XCO~factor.  Column 3: the CO (3-2) to CO (1-0) ratio (Sec. \ref{sec:flux_measurements}), where $L'$ is in unit of K km s$^{-1}$ pc$^2$. The errors are inferred from the RMS noise. 
\\$^a$ The molecular outflow mass is estimated from the CO (3-2) flux assuming $L'_{\mathrm{CO (3-2)}}/L'_{\mathrm{CO (1-0)}} =1$. 
\\$^b$  $3~\sigma$ lower-limit. 
\\$^c$ This is a conservative upper-limit inferred from the CO (1-0) 3 $\sigma$ detection limit and the \XCO~uncertainty is taken into account. 
}
\end{deluxetable}

\subsection{Molecular Gas Mass and Star Formation Rate}
\label{sec:gas_mass}

From the measured CO (1-0) luminosity we can estimate the molecular gas
 mass \citep[e.g.,][]{2013ARA&A..51..207B}. The masses are listed in Table \ref{tab:gas_prop}. 
The CO-to-H$_2$ conversion factor \XCO~
is defined as
\begin{align}
X_{\mathrm{CO}} = N_{\mathrm{mol}}/W(\mathrm{CO (1-0)}),
\end{align}
where $N_{\mathrm{mol}}$ is the molecular column density in units
of cm$^{-2}$, and $W(\mathrm{CO (1-0)}$) is the CO (1-0) brightness in units of K $\mathrm{\kms}$. Although \XCO\ is
roughly constant in normal Galactic
molecular clouds \citep{1987ApJ...319..730S}, it is found to be lower in ULIRG/starburst
galaxies by a factor of 2-5 with large scatter
\citep{1998ApJ...507..615D}. Likely because the molecular clouds blend together due to tidal forces in the dense environment of the ULIRG nucleus, the CO emission emerges from a diffuse volume-filling medium rather than discrete self-gravitating molecular clouds. 

 SDSS J1356+1026 is identified as
a ULIRG and has disturbed gas dynamics, so we adopt an \XCO~value for ULIRGs 
 \XCO = $0.4 \times 10^{20}$ cm$^{-2}$ (K~\kms)$^{-1}$ with an error of 0.5 dex  \citep[][]{1998ApJ...507..615D,2013ARA&A..51..207B}. This corresponds to \alphaCO = 0.86 \msun~(K~kms~pc$^{2}$)$^{-1}$, matching the value used in other ULIRG/AGN CO studies \citep[e.g.,][]{2014A&A...562A..21C}.
This factor of three uncertainty in the \XCO~factor dominates the errors in the mass estimates. We find a total
molecular mass of $9^{+19}_{-6}\times10^{8}~\msun$. Half of the molecular mass is in
the extended W arm ($5^{+11}_{-3}\times10^{8}~\msun$), while the N nucleus
shares a third ($3^{+6}_{-2}\times10^{8}~\msun$). As CO (1-0) is only marginally detected at the S nucleus, a conservative mass limit of $<3.8\times10^{8} \msun$ is estimated from the 3 $\sigma$ detection limit and the uncertainty in \XCO.

Since molecular gas is the fuel for star formation, we can estimate the star formation that can be supported by the molecular content in SDSS J1356+1026.  Using the Schmidt-Kennicutt law \citep{KennicuttJr:1998id}, we find the star formation rate at the N nucleus to be SFR=1.2 \msun~yr$^{-1}$ ($\pm$ 0.5 dex) 
assuming that all of the molecular gas at the N nucleus is distributed in a disk with a radius of 300 pc, half of the beam deconvolved source FWHM in CO (3-2). 
As the morphology of the molecular gas in the W arm and the S nucleus are uncertain, their SFR is poorly constrained. 
A conservatively high estimate of the entire system can be calculated by putting all the molecular gas, including that in the diffuse W arm, in a disk with a radius of 300 pc, which gives 5 \msun~yr$^{-1}$. 
Taking the factor of three \XCO~uncertainty into account, an absolute upper-limit is placed at 16 \msun~yr$^{-1}$. 
However, the star formation rate inferred from the CO luminosity is indirect. 
Whether the assumed \XCO~factor and/or the Schmidt-Kennicutt law apply in this environment is uncertain. In the following section, we constrain the star formation rate directly from the infrared spectral energy distribution (SED) decomposition.

\subsection{SED, Star Formation Rate and AGN Bolometric Luminosity}
\label{sec:SED}
We use the infrared spectral energy distribution (SED) decomposition to constrain the star formation rate (Fig. \ref{fig:SED}, Table \ref{tab:SED}). 
It is usually thought that the dust close to the AGN is heated to much higher temperature than the diffuse dust heated by starlight. The difference in the dust temperatures is at the core of the SED-fitting method to distinguish between AGN-dominated and star formation-dominated sources.
\citet{2012ApJ...746...86G} compiled a UV to radio spectral energy distribution (SED) of SDSS J1356+1026 making use of the 2MASS \citep{Skrutskie:2006hl}, \emph{IRAS} \citep{1984ApJ...278L...1N}, FIRST \citep{1995ApJ...450..559B}, and NVSS \citep{Condon:1998kn} data. 
We further include data from the Wide-field Infrared Survey Explorer \emph{WISE} \citep[][]{2010AJ....140.1868W} 
and the \emph{Herschel} Space Observatory \citep{Pilbratt:2010en} PACS \citep{Poglitsch:2010bm} and SPIRE \citep{Griffin:2010hz} data from Petric et al. in prep., which extends to 500 \micron~to cover the Rayleigh-Jeans tail of the dust emission. 

Following the decomposition procedures of \citet{2012ApJ...759..139K} we fit a two-temperature modified blackbody to the seven-point SED. This SED of 22 \micron~\emph{WISE}, 60 \micron~\emph{IRAS}, and 70-350 \micron~\emph{Herschel} is sensitive to the temperature of the dust emission.  
There are three free parameters in the model: the temperature $T_{\rm{warm}}$ and luminosity $L_{\rm{warm}}$ of the warm component and the luminosity of the cold component $L_{\rm{cold}}$. The temperature of the cold dust is fixed at $T_{\rm{cold}}=35$ K, as found in a wide range of systems by \citet{2012ApJ...759..139K}. 
Even if we fit the temperature of the cold component we recover $T_{\rm{cold}}=35$ K.
The emissivity of the dust is assumed to be a power-law function of frequency with index $\beta=1.5$. All photometry errors are assumed to be 10\%, which is representative of the systematic errors. 
The best-fit model (black line, Fig. \ref{fig:SED}) matches the data well with a reduced $\chi^2$ of 1. 
The best-fit temperature of the warm component is high $T_{\rm{warm}}=81$ K, and the luminosity is $L_{\rm{warm}}=2.6\times10^{45}$ \ergs, much higher than that of the cold component $L_{\rm{cold}}=4.3\times10^{44}$ \ergs. 
 
The luminosity-weighted temperature $T_{\rm{eff}}=75$ K is higher than that seen in \citet{2012ApJ...759..139K} even in those of their objects that have an AGN dominated SED ($T_{\rm{eff}}=65$ K). 
Therefore, as the warm component is consistent with being AGN-heated, the AGN is the major heating source for the bulk of the dust-emission. 
However, star-formation cannot be ruled out as the heating source for the 35 K cold component. 
We therefore use the luminosity of this 35 K component as an upper limit on the luminosity of dust emission associated with star formation, and use calibrations from \citet{Bell:2003bj} to calculate an upper limit on the star formation rate of 21 \msun~yr$^{-1}$. 
 If the SFR were much higher, at the same $T_{\rm{cold}}$, the dust emission would exceed the measured 250 \micron~to 350 \micron~flux. 
This SFR upper-limit is insensitive to the temperature and luminosity of the warm component because the 250 \micron~and 350 \micron~ luminosities are dominated by the cold component.
\citet{_2014arXiv1403.3086H} estimated a higher star formation rate of $63^{+7}_{-17}$ \msun~yr$^{-1}$, inconsistent with our limit. This high SFR is inferred from the \emph{WISE} and \emph{IRAS} data alone covering 4.6 \micron~to 100 \micron. In this case, the \emph{Herschel} photometry at 250 \micron~and 350 \micron~are critical for the SED decomposition and SFR estimation. 

We plot our measured 100 GHz flux (Sec. \ref{sec:100GHz}) in the SED (right panel of Fig. \ref{fig:SED}). 
The flux at 100 GHz is much higher than the extrapolated Rayleigh-Jeans tail of the dust emission, and thus must have a different origin. We find a spectral slope of $\alpha = -0.86$ ($F_{\rm{\nu}}\propto \nu^\alpha$) between the 100 GHz and the 1.4 GHz fluxes, similar to the typical spectral indices $\alpha \approx -0.7$ seen from synchrotron emission for radio-loud AGN \citep{Zakamska:2004fy}. 

Mid-infrared (5-25 \micron) luminosity has also been used as a bolometric indicator for the AGN. 
We use the \emph{WISE} photometry to infer \lbol~following similar procedures as \citet{2013MNRAS.436.2576L}. As pointed out by \citet{2013MNRAS.436.2576L}, type 2 quasars have significantly redder mid-infrared colors than type 1 unobscured quasars, possibly due to dust reddening. 
This red color is also seen in SDSS J1356+1026. Therefore, to avoid attenuation from the dust, we choose the reddest \emph{WISE} band at 22 \micron~ (19.5 \micron~ rest-frame) to apply the bolometric correction of $\lbol=\nu L_{\nu}\times (11\pm5)$ from \citet{2006ApJS..166..470R}, which gives $\lbol \approx 1.1\pm0.5\times 10^{46}$ \ergs. 
However, this \lbol~might still be underestimated due to the strong extinction. 
We further use the 30 \micron~luminosity density extrapolated from the 3.4, 4.6, 12, and 22 \micron~data assuming a power-law spectral shape to set a conservative upper-limit of $\lbol < 2.3 \times 10^{46}$ \ergs~with a bolometric correction factor of 13 \citep{2013MNRAS.436.2576L}. 
The \lbol~estimate from the mid-infrared data is similar to that inferred from the \oiii~luminosity, \lbol $\approx 10^{46}$ \ergs \citep{2012ApJ...746...86G}, using a \lbol-\loiii~relation \citep{Liu:2009gr} calibrated with type 1 AGN.

\begin{deluxetable}{cccc}[h!]
\tablecolumns{5}
\tabletypesize{\scriptsize}
\tablewidth{0pc}
\tablecaption{Spectral Energy Distribution \label{tab:SED}}
\tablehead{
\colhead{Band} & \colhead{$\lambda_{\rm{rest}}$} & \colhead{$\nu_{\rm{rest}}$} & \colhead{$\nu L_{\nu}$}\\
\colhead{} & \colhead{(\micron)} & \colhead{(Hz)} & \colhead{($10^{44}$ \ergs)}\\
\colhead{(1)} & \colhead{(2)} & \colhead{(3)} & \colhead{(4)}}
\startdata
\emph{WISE} W1 3.4 \micron &	3.0	&	$9.9\times10^{13}$	&	0.80	\\% $\pm$ 0.08 \\
\emph{WISE} W2 4.6 \micron &	4.1	&	$7.3\times10^{13}$	&	1.21	\\% $\pm$ 0.12 \\
\emph{WISE} W3 12 \micron &	11	&	$2.8\times10^{13}$	&	2.72	\\% $\pm$ 0.27 \\
\emph{WISE} W4  22 \micron$^{a}$ &	20	&	$1.5\times10^{13}$	&	10.5	\\% $\pm$ 1.1	\\
\emph{IRAS}	 60 \micron$^{a}$ &	53	&	$5.6\times10^{12}$	&	16.1	\\% $\pm$ 1.6	 \\
\emph{Herschel} PACS	70 \micron	$^{a}$ &	62	&	$4.8\times10^{12}$	&	13.5	\\% $\pm$ 0.19	\\
\emph{Herschel} PACS	100 \micron$^{a}$	&	89	&	$3.4\times10^{12}$	&	7.50	\\% $\pm$ 0.15	\\
\emph{Herschel} PACS	160 \micron$^{a}$	&	142	&	$2.1\times10^{12}$	&	2.18	\\% $\pm$ 0.09	\\
\emph{Herschel} SPIRE 250 \micron$^{a}$	&	223	&	$1.3\times10^{12}$	&	0.59	\\% $\pm$ 0.03	\\
\emph{Herschel} SPIRE 350 \micron$^{a}$	&	312	&	$9.6\times10^{11}$	&	0.16	\\% $\pm$ 0.02	\\
\emph{Herschel} SPIRE 500 \micron	&	445	&	$6.7\times10^{11}$	&	$<$0.08		\\
ALMA 100 GHz	 &	$2.67\times10^{3}$	&	$1.1\times10^{11}$	&	$6.10 \times10^{-4}$	
\enddata
\tablecomments{The rest-frame mid-to-far infrared spectral energy distribution of SDSS J1356 +1026 (Sec. \ref{sec:SED}, Fig. \ref{fig:SED}). Column 1: the telescope and band. Column 2: the rest frame wavelength. Column 3: the rest frame frequency. Column 4: the frequency times the luminosity density in the rest frame. The errors in the infrared data (\emph{WISE}, \emph{IRAS}, and \emph{Herschel}) are assumed to be 10\%. \\
$^{a}$ The seven data points from the \emph{WISE} 22\micron~to \emph{Herschel} SPIRE 350 \micron~are used for the FIR SED decomposition fitting.
}
\end{deluxetable}

\begin{figure*}[]
	\centering
	\includegraphics[bb = 70 260 530 640,clip, scale=1]{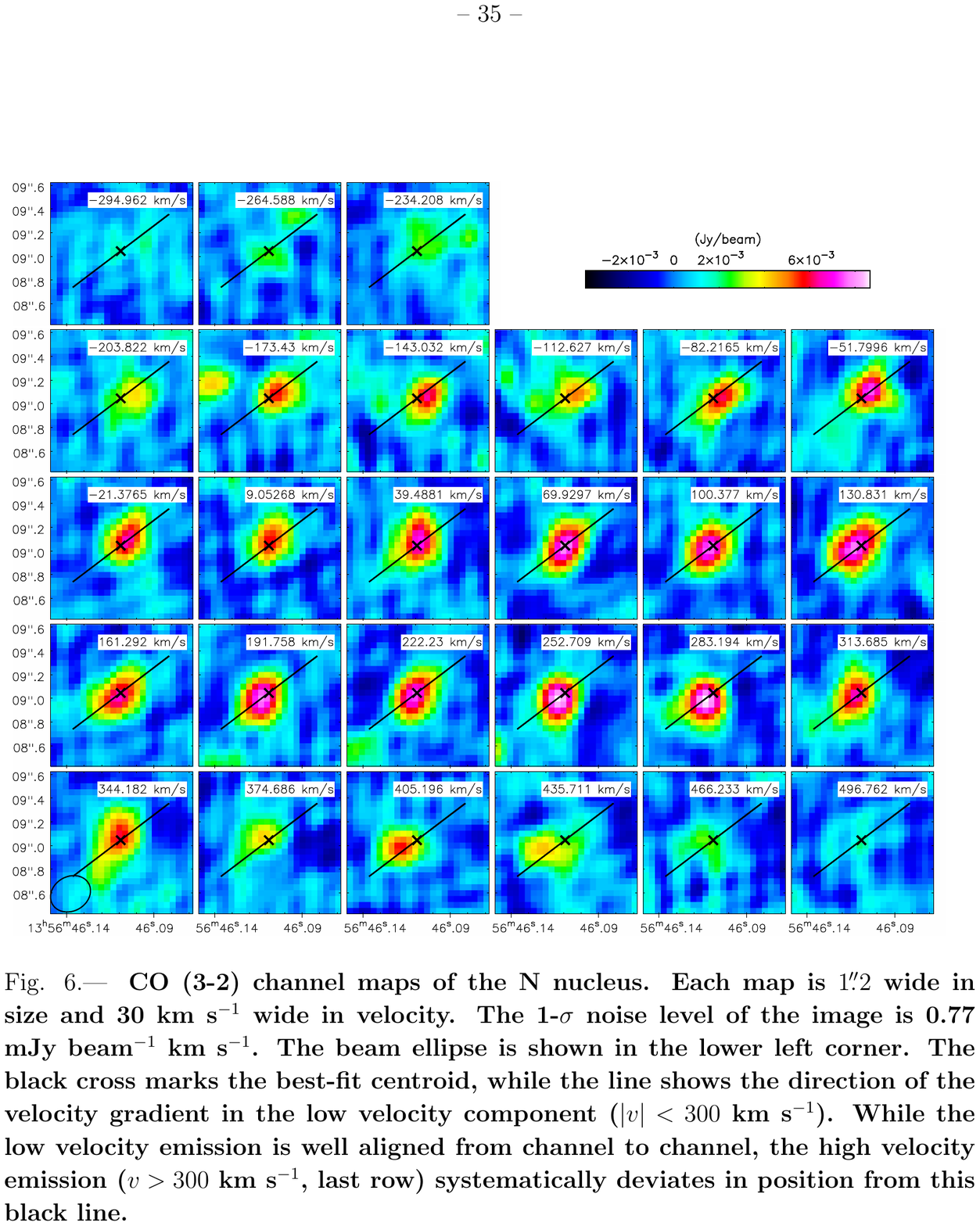}
	 \caption{ CO (3-2) channel maps of the N nucleus. Each map is $1\farcs2$ wide in size and 30 \kms~wide in velocity. The 1-$\sigma$ noise level of the image is 0.77 mJy beam$^{-1}$ \kms. The beam ellipse is shown in the lower left corner. The black cross marks the best-fit centroid, while the line shows the direction of the velocity gradient in the low velocity component ($|v| < 300$ \kms). While the low velocity emission is well aligned from channel to channel, the high velocity emission ($v>300$ \kms, last row) systematically deviates in position from this black line. 
	}
        \label{fig:channel_N}
\end{figure*}

\section{Results}
\label{sec:results}

\subsection{The N nucleus}
\label{sec:mol_OF}
There is strong CO (1-0) and CO (3-2) emission at the N nucleus with a wide velocity distribution from $-$300 \kms~to 500 \kms~(Fig. \ref{fig:spec}). 
From the CO (1-0) flux, the molecular gas mass is estimated to be $3^{+6}_{-2} \times10^8$ \msun, which can fuel star formation at an approximate rate of $\sim$ 1 \msun~yr$^{-1}$, if the molecular gas is in a disk of radius 300 pc. The N nucleus has a CO line ratio $L'_{\mathrm{CO (3-2)}}/L'_{\mathrm{CO (1-0)}} = 1.18\pm0.73$. 
As various factors affect the line ratio, including the gas temperature and density,
we are unable to infer the physical conditions of the gas from this one ratio.
 Instead, we empirically compare our measured line ratio to other objects using the compilation in \citet{Carilli:2013il}, including quasars ($L'_{\mathrm{CO (3-2)}}/L'_{\mathrm{CO (1-0)}}\approx$ 1.0), sub-millimeter galaxies ($\sim 0.7$), normal star-forming galaxies \citep[$\sim$0.4-0.6,][]{2013ApJ...763...64B}, and the Milky Way ($\sim$0.3). The N nucleus has a relatively high line ratio similar to that of a quasar.

The CO (3-2) emission is marginally resolved (FWHM = $0\farcs45 >$ Beam = $0\farcs35$) with a decomposed FWHM of $0.6\pm0.1$ kpc. 
This spatial resolution allows us to analyze the molecular kinematics at the N nucleus. The channel maps (Fig. \ref{fig:channel_N}) show that the emission moves toward the south-east as the velocity increases until about 300 \kms. 
Above 300 \kms, there is high velocity emission extending to 500 \kms, but the position does not align with the low velocity gas. 

In each channel we fit the emission with a 2-D Gaussian profile, and derive the best fit x and y positions (Fig. \ref{fig:channel_N}). We fit a linear model $\textbf{\emph{x}}=\textbf{\emph{a}}v+\textbf{\emph{b}}$ to the best fit positions in the velocity range of $\pm300$ \kms~to yield a PA $=127 \degree$. 
We then extract a position-velocity diagram (hereafter $pv$-diagram, Fig. \ref{fig:pv}) along this PA. 
The $pv$-diagram is extracted over a width of $0\farcs5$, which encloses $> 95\%$ of the flux. 
It can be seen in the $pv$-diagram that the emission within $\pm300$ \kms~roughly follows a linear position-velocity relation that is symmetric about the systemic velocity and has a best-fit gradient of 1077 \kms~kpc$^{-1}$. The stellar velocity dispersion at the N
nucleus (within a $1\arcsec$ aperture) is $\sigma_* = 206 \pm 36$
\kms~ \citep{Greene:2009kt}, 
which would correspond to a circular velocity of $V_c = \sqrt{2}\sigma_* = 291 \pm 50$ \kms~for an isothermal sphere. Therefore, the linear velocity gradient within $\pm300$ \kms~is consistent with a rotating disk.  
This velocity gradient corresponds to a constant dynamical mass density of $\rho_{\rm{dyn}} \approx 63$ \msun\ pc$^{-3}$, which gives a dynamical mass within a radius of 0.25 kpc (0.3 kpc) of $M_{\rm{dyn}} \approx 4\times10^9$ \msun\ ($7\times10^9$ \msun). The molecular gas mass of $M_{\rm{mol}} = 3^{+6}_{-2} \times10^8$ \msun~inferred from the CO emission accounts for $1-20 \%$~of the dynamical
mass in the N nucleus.

Beyond this rotating component, the red-shifted high velocity feature
 ($300 < v < 500$ \kms) deviates from the linear velocity gradient and 
 has a velocity too high to be explained by rotation alone. 
The bottom of Figure \ref{fig:channel_N} and the left panel of Figure 
\ref{fig:pv} show that the position of this feature has a slight offset ($0\farcs02$) to the north-east from the disk plane. 
There is no counterpart on the blue-shifted side. 
The CO (3-2) emission is detected with 9~$\sigma$ significance ($0.99\pm0.10$ Jy \kms), while there is only a marginal detection $0.06 \pm 0.08$ Jy \kms~for CO (1-0). 
Therefore, the line ratio $L'_{\mathrm{CO (3-2)}}/L'_{\mathrm{CO(1-0)}}=1.8\pm2.5$ is poorly constrained, but hints at a high excitation level. 
The 3 $\sigma$ detection limit of CO (1-0) places a lower limit on the line ratio $L'_{\mathrm{CO (3-2)}}/L'_{\mathrm{CO(1-0)}} > 0.45$.
We use the CO (3-2) flux to infer a molecular mass $M_{\mathrm{mol}} =7^{+15}_{-5}\times10^7$~\msun, assuming a line ratio of $L'_{\mathrm{CO (3-2)}}/L'_{\mathrm{CO(1-0)}}=1$ as observed in the bulk of the N nucleus as well as in
other AGN. This high velocity component accounts for a significant fraction ($\sim 20\%$) of the molecular mass in the N nucleus, under the assumption of constant line ratio and \XCO~factor. 
We discuss the interpretation of this component in Section \ref{sec:discussion_molOF}.
\\

\subsection{The W arm}

The extended emission to the west of the N nucleus is 2\arcsec~(4 kpc) long, and has a blue-shifted narrow line at $-150$ \kms~with a FWHM of 200 \kms~(Fig. \ref{fig:spec}). The CO (1-0) flux is $0.82\pm0.34$ Jy \kms~once the overlap with the N nucleus is accounted for, and the CO (3-2) flux is $2.55\pm 0.40$ Jy \kms. 
Adopting a ULIRG \XCO~ratio, the W arm contains $5^{+11}_{-3} \times10^8$ \msun~ of molecular gas, about half of the total mass in the system. The actual mass could be even higher, up to $2.5\times10^9$ \msun, if the molecular gas is in self-gravitating clouds and the higher Milky Way \XCO~ratio applies. 
It has a moderate line ratio of $L'_{\mathrm{CO (3-2)}}/L'_{\mathrm{CO (1-0)}} = 0.35\pm0.20$, which is lower than the N nucleus and is similar to star forming galaxies \citep[0.4-0.6,][]{2013ApJ...763...64B,Carilli:2013il}. 

The W arm is likely a tidal feature (Sec. \ref{sec:origin}). It is only slightly offset from the position of an extended stellar plume to the north west of the N nucleus (see the CO (1-0) map in Figure \ref{fig:moment0}).  
Decoupling between stellar and gas components in merging systems is not only predicted by simulations \citep[e.g.,][]{1996ApJ...471..115B} but also commonly seen in observations of tidal tails \citep[e.g.,][]{2000AJ....119.1130H}, as the gas is subject to dissipational processes that can affect its trajectory. 
Furthermore, the cool and coherent kinematics of the W arm, as seen from its narrow line width (Fig. \ref{fig:spec}), across its full length of 4 kpc, makes it hard to explain by either outflow or inflow.

\subsection{The S nucleus}
At the location of the southern nucleus identified from the \emph{HST} images, there is only a $\sim2~\sigma$ CO (1-0) detection of $0.13\pm0.06$ Jy \kms, and no detection for CO (3-2). The 3 $\sigma$ detection limit of CO (1-0) translates into a molecular gas mass of $1.2\times10^8$ \msun, about 10\% of the total molecular gas mass in the system. 
Taking the uncertainty in the \XCO~factor into account, we can place an upper-limit of $M_{\rm{mol}}<3.8\times10^8$ \msun. 
Therefore, the S nucleus is not a major reservoir of molecules in SDSS J1356+1026. 
Although the sensitivity is not adequate to constrain the detailed line profiles, the CO (1-0) marginal detection suggests a narrow blue-shifted line centered on $-30$ \kms~(Fig. \ref{fig:spec}). The much narrower CO (1-0) line width compared to the N nucleus can result from the shallower gravitational potential or the absence of gas outflow/inflow in the S galaxy. 
The  $L'_{\mathrm{CO (3-2)}}/L'_{\mathrm{CO (1-0)}}$ line ratio is constrained to be $0.10\pm0.11$, much lower than the N nucleus and the W arm.

\subsection{Molecular Counterpart to the Ionized Bubble}
\label{sec:mol_counterpart}

As we describe in Section \ref{sec:intro_bubble}, SDSS J1356+1026 contains an extended ionized outflow discovered by \citet{2012ApJ...746...86G}. 
However, there is no detection of molecular gas at the corresponding location. 
 We identify the region of ionized line emission from the elongated luminous structure in the \emph{HST} F814W image (green region in Fig. \ref{fig:moment0}). 
The integrated CO (1-0) flux of $-32\pm102$ mJy \kms~places a
molecular gas mass upper-limit (95$\%$ c.l.) of $M_{\rm{mol}} < 8.6\times 10^7$ \msun, while \citet{2012ApJ...746...86G}
give a lower-limit on the ionized gas mass $M_{\rm{ion}} >
5\times10^7$~\msun. Therefore, the molecular gas does not dominate the mass of the extended ionized outflow. 

Despite the non-detection of CO in the ionized outflow, we note a
tentative detection of a double-peaked CO (3-2) emission line at an
optically bright spot in the \emph{HST} F814W image about 4\arcsec
~south of the N nucleus where the \oiii~ expanding-shell signature
starts to emerge (the ``bubble base'' shown in magenta in
Fig. \ref{fig:moment0}). The two CO (3-2) peaks, each having 3 to
4~$\sigma$ significance, are at $-600$ and $+400$ \kms\ respectively.
Despite the intriguing location and velocity of this component, its confirmation requires further observations.

\begin{figure*}[]
        \centering
        \hbox{
	\includegraphics[bb=83 300 500 800,clip, scale=0.49]{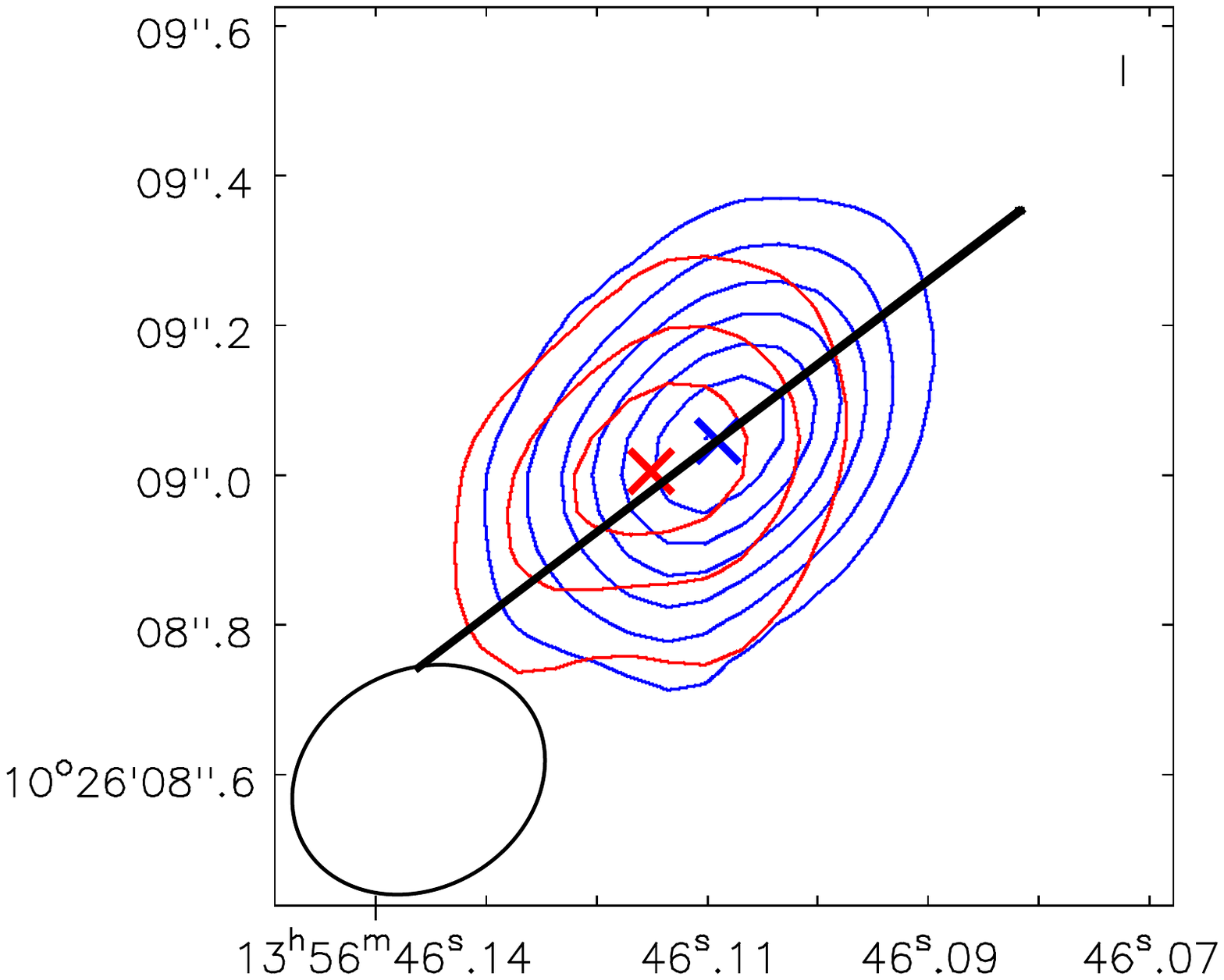}
	\includegraphics[bb=80 40 950 800,clip, scale=0.3]{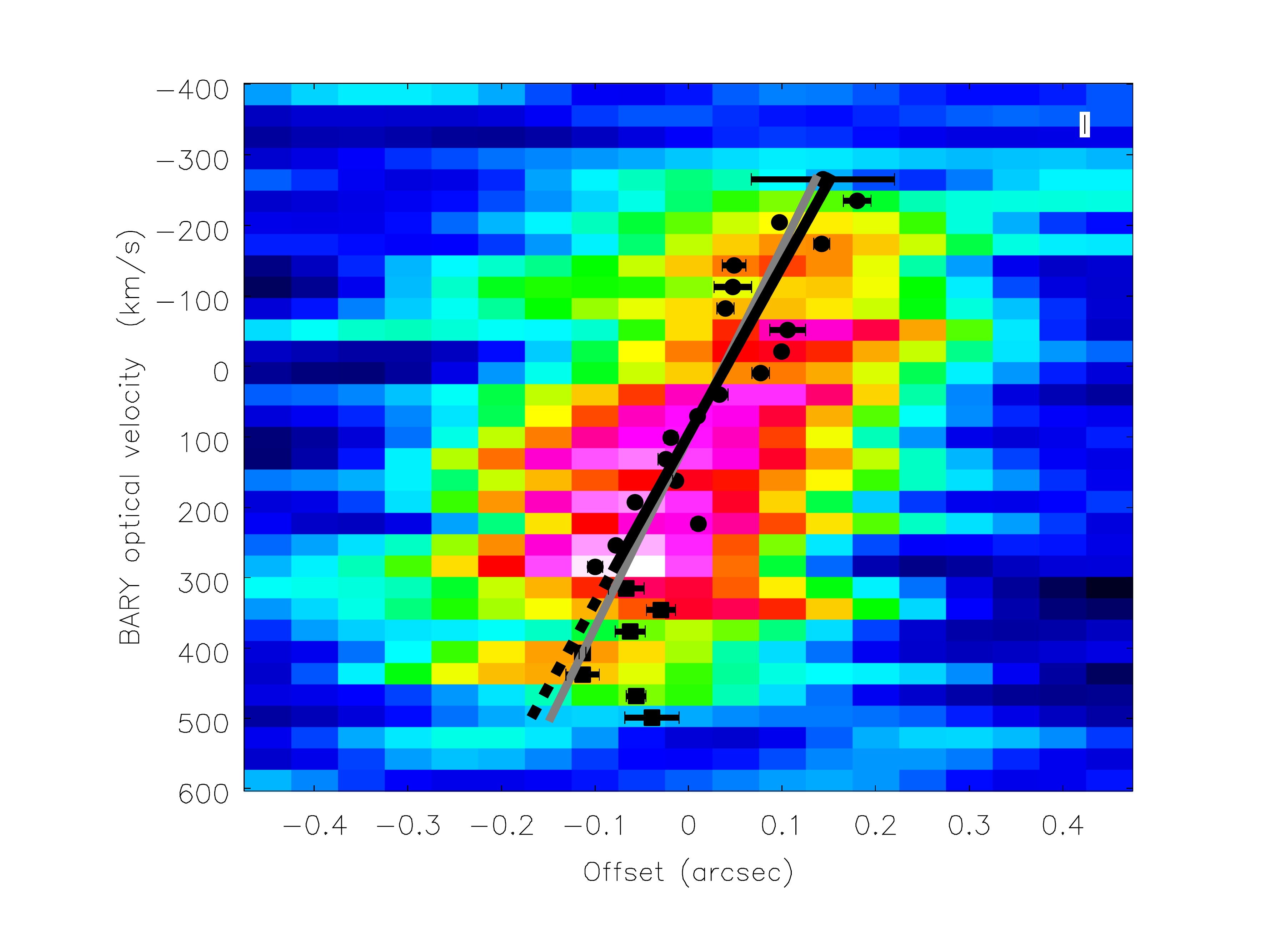}
	}
        \caption{
        Left: CO (3-2) N nucleus moment-0 map of the low velocity component $|v| < 300$ \kms~(blue contours) and the high velocity component $v>300$ \kms~(red contours). The contours have a step size of 3 $\sigma$ (0.48 and 0.22 Jy beam$^{-1}$ \kms, respectively) between each level. The crosses show the best-fit Gaussian centroids of the two components. The beam ellipse is shown in the lower left corner. The direction of the velocity gradient of the low velocity component (P.A. $=127 \degree$) is indicated by the blue line. We extract a {\it pv}-diagram along this blue line and show it on the right panel. It is over a width of $0\farcs55$, such that it contains $\sim 95 \%$ of the flux. The offset on the horizontal axis is with respect to the best-fit Gaussian centroid of the low velocity component (blue cross on the left panel). The black points represents the centroid position at each velocity channel, fitted by Gaussian intensity profile. The black line shows the best-fit linear velocity gradient (1077 \kms~kpc$^{-1}$) of the low velocity component ($|v| < 300$ \kms, black circles) that is consistent with a rotating disk, and the grey line is fitted to all the velocity channels ($-300 < v < 500$ \kms). The high velocity points (black squares, $v>300$ \kms) systematically deviate from this velocity gradient, and have velocities too high to be consistent with rotation. }
        \label{fig:pv}
\end{figure*}

\section{Discussion}
\label{sec:discussion}

\subsection{Origin of the Molecular Gas}
\label{sec:origin}
We first discuss the origin and evolution of the molecular gas in the system. 
A significant amount of molecular gas is detected in the N nucleus and the W arm with a total mass of $9^{+19}_{-6}\times10^8$ \msun~ (Section \ref{sec:analysis}). 
However, the stellar absorption spectrum of the N nucleus is dominated by an old stellar population \citep{Greene:2009kt}, and based on the velocity dispersion we suspect that the galaxy started as an early-type with little cold gas. 
For local elliptical/S0 galaxies like SDSS J1356+1026 with $\sigma_* > 200$ \kms,  \citet{2011MNRAS.414..940Y} find that only one out of thirty galaxies has a molecular gas mass as high as $8\times10^8$ \msun, while most of the ellipticals are below the CO detection limit of $M_{\rm{mol}}<10^8$ \msun. 
Therefore, the exceptionally high molecular content in the N galaxy was likely accreted externally, perhaps from the S galaxy. The current optical spectroscopy is not adequate to identify the stellar population of the S galaxy. But if it was a late-type galaxy, with a stellar mass of about $M_{*}\approx3\times10^{10}$ \msun, it could bring in plenty of molecular gas, as the typical molecular to stellar mass ratio is about 5\% for local spirals \citep{2009AJ....137.4670L, 2013ApJ...768..132B}. 
Gas transfer from disk companions to elliptical galaxies has been observed \citep{1987IAUS..127..109S}. Furthermore, some early-type galaxies have cold gas components that are kinematically decoupled from the stellar components, suggesting an external origin of the cold gas \citep{1987IAUS..127..145K, 2013MNRAS.429..534D}. 
Therefore, a primary elliptical/satellite disk galaxy merger is a likely scenario to explain the molecular gas content in SDSS J1356+1026.  

In simulations, when the gas collides during a merger, its orbital energy and angular momentum are dissipated, and thus the gas tunnels into the center 
to form a nuclear disk and triggers AGN or star formation activity \citep[e.g.,][]{1989Natur.340..687H, 2002MNRAS.333..481B}. 
In particular, if it is a primary elliptical and satellite disk galaxy merger, in some cases the gas in the disk galaxy is disrupted and sinks into 
 the center of the primary elliptical \citep{1993ApJ...405..142W}. 
These nuclear gas disks, growing from infalling tidal tails, usually contain about half of the cool gas in the system and have peculiar kinematics relative to the stars. 
While these simulations do not treat the molecular gas phase explicitly, we expect the nuclear disks to have a large molecular fraction because of the dense environment. 
Observationally, molecular nuclear gas disks and rings
are indeed common in infrared luminous mergers \citep{1998ApJ...507..615D}. These compact nuclear disks have radii $\sim 0.5$ kpc, as opposed to $\sim1$ kpc in normal ellipticals \citep{2013MNRAS.429..534D}, and can be as massive as $10^9$ \msun, similar to the compact nuclear disk of SDSS J1356+1026. 
The energetic nuclear activity of SDSS J1356+1026 could be a direct result of the nuclear inflow induced by the interaction between the two merging galaxies \citep{1996ApJ...471..115B}.

Observationally, 
although extended molecular tails can survive after being stripped from the galaxies \citep[e.g., through ram pressure stripping,][]{2012A&A...540A.112D, _2014arXiv1403.2328J}, it is not common to find a significant amount of molecular gas in tidal tails.
\citet{2001A&A...372L..29A} found $4\%$ ($M_{\rm{mol}}\approx 8\times10^7$ \msun) of the molecular gas along a 10 kpc long optical tidal tail in NGC 4194, a elliptical/disk merger remnant. 
The ULIRG merger Mrk 273 also shows a 5 kpc extended CO streamer with mass $M_{\rm{mol}}\approx 1\times10^8$ \msun, accounting for a few percent of the CO flux \citep{1998ApJ...507..615D}. One of the nuclei of Mrk 273 has no molecular gas, so it was possibly a gas poor galaxy or the gas was depleted by transfer \citep{1995ApJ...451L..45Y}. 
However, none of these examples contain as much molecular gas as in the W arm of SDSS J1356+1026 ($5\times10^8$ \msun, half of the total molecular mass). 

Both SDSS J1356+1026 and NGC 4194 are probably elliptical/disk minor mergers. We suspect that the formation of molecular tidal features may require particular progenitor properties and orbital configurations. For example, in the primary elliptical/satellite disk merger, the elliptical with its concentrated potential may disrupt the molecular disk in the satellite galaxy. At the same time, the absence of gas in the elliptical can reduce the shock dissociation of molecular gas during the encounter and thus preserve gas in the molecular phase in tidal features.

\subsection{High Velocity Molecular Gas at the N Nucleus}
\label{sec:discussion_molOF}

In Section \ref{sec:mol_OF}, we find a high velocity CO (3-2) feature at the N nucleus with a radius of $\sim$ 300 pc. This feature ranges in velocity from 300 to 500 \kms, which is too high to be part of the rotating disk. 
Its mass is estimated to be $M \approx 7\times10^7$~\msun, a quarter of the mass at the N nucleus.  
Also, it has complex kinematics that deviate from rotation and has a slight offset with respect to the disk plane (Fig. \ref{fig:channel_N}, \ref{fig:pv}).  
There are various possible interpretations for this high velocity component. It could be inflowing or outflowing, or could result from tidal forces or feedback mechanisms. We discuss two of the most plausible scenarios.

First, the high velocity feature may represent material accreting onto the disk from an external source, perhaps the W tidal arm.  
The projected angular momentum of the two features are parallel to each other, although they lie on opposite sides of the disk.  Furthermore, the angular momentum of the high velocity gas is smaller than the W arm, as expected for gas that is accreting onto a nuclear disk.

The second possibility is a molecular outflow. \emph{Herschel}  OH line observations have found molecular outflow to be a common phenomenon in ULIRG/AGN like SDSS J1356+1026 \citep{2013ApJ...776...27V}. 
The high velocity feature seen in SDSS J1356+1026 meets
one of the outflow criterion used by the CO interferometry study by \citet{2014A&A...562A..21C}, in that it has a velocity above 300 \kms~and deviates from the rotation pattern, making it a likely outflow candidate. 
However, CO confirmed molecular outflows in luminous ULIRG/AGN are typically more massive and larger in size. 
The six molecular outflows in luminous ULIRG/AGN (\lbol $>10^{44.5}$) discussed by \citet{2014A&A...562A..21C}, including IRAS F08572+3915, IRAS F10565+2448, IRAS 23365+3604, Mrk 273, Mrk 231, and NGC6240, have molecular outflow masses $M \approx 3\times10^8$, sizes $r = $0.6-1.2 kpc, and velocities $v =  $400-1200 \kms. 
This discrepancy in size could be partly due to the high angular resolution of the ALMA observations. Structures of size 300 pc may not have been resolved by previous CO studies. 
The fact that this compact size is measured by CO (3-2), a denser tracer than CO (1-0), may also contribute to the discrepancy.
In fact, very compact molecular outflows on the scale of $\sim$100 pc have been discovered by OH observations combined with radiative-transfer modeling \citep{2011ApJ...733L..16S, GonzalezAlfonso:2013iq}. 

With the current observations we are unable to completely confirm or rule out either scenario.  Given the ubiquity of nuclear outflows observed with \emph{Herschel} on similar scales, in what follows we elaborate further on the outflow scenario and discuss the implications of the possible compact molecular outflow in this source.

\subsubsection{Properties of the Molecular Outflow}

We adopt a size of $r = 0.3$ kpc, velocity $v= 500$ \kms, and mass $M \approx 7\times10^7$~\msun~for the molecular outflow. These quantities are only order of magnitude estimates. In addition to measurement error, the size and velocity are subject to projection effects. 
There are also potential systematics in the mass estimate. 
First, the $X_{\rm{CO}}$ factor in the context of an outflow is uncertain. The lower ULIRG $X_{\rm{CO}}$ factor should already account for (at least part of) the fact that the molecular gas in outflows is not in self-gravitating clouds, but little is known about whether the violent kinematics in the outflow would further enhance the CO luminosity and thus reduce the $X_{\rm{CO}}$. 
In principal, if CO (1-0) became optically thin due to very turbulent velocity structure, the $X_{\rm{CO}}$ could further decrease by a factor of 2-3 \citep{2013ARA&A..51..207B}. 
Second, we assume a typical quasar line ratio of $L'_{\mathrm{CO (3-2)}}/L'_{\mathrm{CO(1-0)}}=1$, but we have only a loose constraint from the observations. Potentially, the shocks in the outflow could induce a different temperature profile and thus affect the line ratio, although this effect is not seen in Mrk 231, where the excitation ratios of the molecular outflow and the bulk of the molecular gas are similar \citep{2012A&A...543A..99C}.

We estimate the time scale, outflow rate, momentum, and energetics of the outflow. 
The dynamical time is very short, $t_{\rm{dyn}}=r/v\approx 0.6$ Myr, reflecting the compact size. 
The mass outflow rate depends on the geometry of the outflow, which cannot be constrained with the current data. We consider two limiting cases. If the outflow is a single burst and the material is distributed in a shell (or clump) with a distance $r_{\rm}$ from the center, then the mass outflow rate is $\dot{M}=Mv/r \approx 116$ \msun~yr$^{-1}$. 
If the outflow is continuous and volume filling inside a sphere or cone with a constant velocity, the mass outflow rate is $\dot{M}=3Mv/r \approx 350$ \msun~yr$^{-1}$, a factor of three higher due to the ratio between the surface area and the volume of a sphere. 
In the following we assume a volume-filling continuous outflow, as in \citet{2012MNRAS.425L..66M} and \citet{2014A&A...562A..21C}. 
The time it takes to deplete the molecular gas in the N nucleus ($\sim 3\times10^8$ \msun) with a constant mass outflow rate is $t_{\rm{dep}}\approx$1 Myr. 
The kinetic energy and kinetic luminosity of
the outflow are $E_{\rm{kin}} \approx 2\times10^{56}$ ergs and
$\dot{E}_{\rm{kin}} \approx 3\times10^{43}$ \ergs,
and the momentum and momentum rate are $p\approx7 \times10^{48}$ g cm
s$^{-1}$ and $\dot{p}\approx1\times10^{36}$ dyne.

\subsubsection{AGN or Star Formation Driven Outflow?}
\label{sec:driving_source}

We now discuss the possible driving mechanism of this compact molecular
outflow.  To examine whether star formation can power the molecular
outflow, we adopt a star formation feedback efficiency of
$\dot{E}_*/SFR = 7\times10^{41}$ \ergs~(\msun~yr$^{-1}$)$^{-1}$
calculated by \citet{1999ApJS..123....3L} using the stellar evolution code
{\it Starburst99} \citep[see also][]{2005ARA&A..43..769V}. The
energy injection rate is dominated by supernovae that occur 40 Myr after
the starburst; at earlier times the rate is
lower. Therefore, to power an outflow with a kinetic luminosity of
$\dot{E}\approx3\times10^{43}$ \ergs, we need a star
formation rate of at least SFR $> 43$~\msun~yr$^{-1}$, much
higher than the star formation rate that can be sustained by the
 molecular gas in the N nucleus (SFR $\approx 1.2$ \msun~yr$^{-1}$), assuming the Schmidt-Kennicutt law. It is also higher than our conservative upper-limits inferred from either the total molecular gas content (SFR $<16$ \msun~yr$^{-1}$ Sec. \ref{sec:gas_mass}) or the far-infrared SED fitting (SFR $<21$ \msun~yr$^{-1}$ Sec. \ref{sec:SED}). 
To fuel a SFR of 43 \msun~yr$^{-1}$ in a disk of radius 300 pc requires a molecular mass of $4\times10^9$ \msun~\citep{KennicuttJr:1998id}, which is 
an order of magnitude higher than the molecular mass in the N nucleus and is equal to its dynamical mass. Therefore, we conclude that there is not enough star formation activity to drive the compact molecular outflow. 

A jet is another possible way to drive the
outflow. As discussed in \citet{2012ApJ...746...86G}, SDSS
J1356+1026 is classified as a radio-quiet quasar, and its radio
emission is not resolved by the FIRST survey
\citep[beam size $5\farcs4$,][]{1995ApJ...450..559B}, which puts a constraint on the size $r < 2$\arcsec~(4 kpc).
The agreement of flux between FIRST and NVSS, which has a larger beam size of 45\arcsec, observations suggests that there is no extended radio emission beyond the 2 \arcsec~scale. 
This rules out a jet as the driver for the extended (20 kpc) ionized outflow. 
 Our 100 GHz measurement (Sec. \ref{sec:100GHz}) further constrains the size of the high frequency radio emission to be no larger than $1\farcs9 \times1\farcs3$ ($4.2\times2.9$ kpc). However, as the high frequency emission may trace only the base of the radio jet, we cannot rule out a compact jet $< 2 \arcsec$. Further observations are required to confirm or rule out a jet as the driving mechanism of the compact molecular outflow.

A radiative driven wind from the obscured quasar at the N nucleus \citep{1995ApJ...451..498M, Proga:2000hm} could also be responsible for driving the molecular outflow. 
Hydrodynamics simulations of AGN wind typically adopt a wind kinetic luminosity of a few percent of \lbol~in order to reproduce the observed black hole mass - stellar velocity dispersion ($M_{\rm{BH}}-\sigma$) and the X-ray luminosity - stellar velocity dispersion ($L_{\rm{X}}-\sigma$) relations \citep[e.g.,][]{DeBuhr:2011iz, 2014MNRAS.442..440C}. 
The AGN bolometric luminosity $\lbol \approx 10^{46}$ \ergs, indirectly inferred from the \oiii~and mid-infrared luminosities (Sec. \ref{sec:SED}), is much higher than the kinetic luminosity of the molecular outflow ($\dot{E}\approx3\times10^{43}$~\ergs) with a ratio of only $\dot{E}/\lbol \sim 0.3\%$. 
The ionized outflow ($\dot{E}\approx10^{44-45}$~\ergs,~\citealt{2012ApJ...746...86G}) has a kinetic luminosity $1-10\%$ of \lbol. 
Therefore, the AGN wind is a feasible scenario in terms of energetics to drive the molecular and ionized outflows.

We find that the momentum rate ($\dot{p} \approx 1\times10^{36}$ dyne) of the molecular outflow is similar to the radiation momentum rate from the AGN ($L_{\rm bol}/c\approx3\times10^{35}$ dyne) with a momentum boost of $\dot{p}/(\lbol/c)\approx$ 3.  
This value is somewhat lower than what is inferred for the luminous ULIRG/AGN that have CO outflows confirmed by \citet{2014A&A...562A..21C}, which typically have a momentum boost of $\dot{M}v/(\lbol/c)\gtrsim$ 20. 
This discrepancy may partly be due to the higher sensitivity of ALMA, which allows us to identify features of lower molecular gas mass. 
The lower momentum boost of the outflow in SDSS J1356+1026 indicates that it is more consistent with being momentum-driven than those found by \citet{2014A&A...562A..21C}. 
\citet{2003ApJ...596L..27K} and \citet{2012ApJ...745L..34Z} argue that outflows on small scales tend to be momentum-conserving, because the hot wind of a luminous AGN is efficiently Compton-cooled at the vicinity of the AGN ($r\lesssim$ 1 kpc). Outflows at larger radius, in contrast, are energy-conserving.  Given that the molecular outflow in SDSS J1356+1026 is more compact than other ULIRG/AGN outflows and shows a lower momentum boost, we are possibly starting to approach the  momentum-driven regime of AGN outflows.

\subsubsection{Comparison with the Ionized Outflow}

SDSS J1356+1026 hosts an extended ionized outflow, 
previously discovered by \citet{2012ApJ...746...86G}. In this
section, we examine the relationship between these two
outflows with their very different sizes, time scales, and molecular
contents by tying them together into an
integrated picture of an AGN wind interacting with galactic gas.

It is intriguing that the molecular outflow is much smaller
($r\approx0.3$ kpc) and is characterized by shorter time scales 
($t_{\rm{dyn}}\approx$ 0.6 Myr, and $t_{\rm{dep}}\approx$ 1 Myr) 
than the extended ionized outflow, which has $r^{\rm{ion}}\approx10$ kpc,
$v^{\rm{ion}}\approx 1000$ \kms, and $t^{\rm{ion}}_{\rm{dyn}}\approx 10$ Myr
\citep{2012ApJ...746...86G}. 
The two outflows are different in morphology - the ionized outflow is
symmetrically bipolar in the north-south direction while the compact molecular outflow has only one component which is redshifted and extends to the S-E direction. 
Furthermore, they have different physical conditions, as the molecular gas dominates the small scale outflow, but is ruled out as the dominant component of the extended outflow (Sec. \ref{sec:mol_counterpart}).  
But the biggest puzzle is the short time scale characteristic of the molecular outflow, which raises concerns about its duty cycle: if the molecular outflow is
indeed a transient phenomenon, how did we capture this rare event at
this exact moment?  

There are two scenarios, continuous and discrete outflows, that could account for the fact that we see outflows on very different scales. 
First, it could be that the
quasar has been active and driving winds for more than $10^7$ years, and
the extended ionized and compact molecular outflows are
two segments of a continuous wind-driven outflow stream. 
They are constituted by gas of different phases possibly due to the differences in the environments, such as density. 
However, sustaining an outflow with a constant outflow rate of $\dot{M}=350$ \msun~yr$^{-1}$ for $10^7$ yr would require a gas mass of $\sim 10^{9.5}$ \msun.
Unless we happened to catch the outflow in the last 10\% of its life time, we would expect to find a comparable quantity of gas reservoir in the system, which we do not see. 
There is only $\sim3\times10^8$ \msun~of molecular gas at the N nucleus, which can fuel the outflow for only $\sim$ 1 Myr. 
Even if the molecular gas in the W arm also fuels the outflow, this outflow would only last for $\sim$ 3 Myr. 
There is no other gas reservoir to replenish the N nucleus; the S nucleus contains little molecular gas, and the extended outflow is unlikely to return once it travels with a velocity of 1000 \kms~to 10 kpc, where the escape velocity is 760 \kms, assuming a singular isothermal density profile with $\sigma_*=206$ \kms ~(eq. 2 of \citealt{2011ApJ...732....9G}). 
Furthermore, we do not find $\sim 10^{9.5}$ \msun~of gas in the outflow, although it could be present in a hotter phase. 
Only $\sim 10^7$ \msun~of outflowing ionized gas is observed, although this number is only a lower limit due to the unknown gas density \citep[][]{2011ApJ...732....9G}. 

Discrete outflow is the second possibility. If the AGN activity is 
episodic \citep[e.g.,][]{2013ApJ...763...60S,2014ApJ...782....9H}, the two outflows may be driven by two separate bursts of the AGN, possibly associated with multiple passages of
the companion (S) galaxy. 
The S galaxy has an orbital period of $\sim 5\times10^7$ yr, assuming a circular orbit with a radius of 2.5 kpc or free falling from a distance of 2.5 kpc. 
In this case, the compact molecular outflow would reflect the most recent ($\sim 10^6$ yr)
burst while the extended ionized outflow is the relic of previous activity ($\sim 10^7$ yr ago).  
This scenario has several advantages over the continuous outflow scenario.  First, we can naturally 
connect the time scales of the outflow to the AGN cycle, which is in
the end controlled by the gas supply. 
Every $\sim 10^{7}$ yr, the passage of the S nucleus triggers a new burst of nuclear activity, which in turn drives a nuclear outburst that depletes the nucleus in $\sim 10^{6}$ yr, thus shutting down further activity. 
 It is therefore not
surprising to see the compact molecular outflow with both a dynamical
time and a depletion time of $\sim 10^{6}$ yr.  Also, it is not a
coincidence that we catch this transient molecular outflow at the right
time. Given that we select this object based on its bright quasar luminosity, its feedback is presumably most active.  Furthermore, it could explain why we see a comparable amount of gas ($10^{7-8}$ \msun) in the compact and ionized outflow, if each
burst releases roughly the same amount of energy and expels the same
amounts of gas.  An outflow driven by episodic AGN activity seems to be a
reasonable possibility that is consistent with the various measured time
and mass scales.

\section{Summary}
The sub-millimeter observations of CO (1-0) and CO (3-2) with ALMA of the luminous obscured quasar SDSS J1356+1026 provide insights into molecular gas dynamics during a merger, AGN feedback on molecular gas, and episodic AGN activity on 10 Myr time scales. We summarize the highlights of our study below. 

SDSS J1356+1026 contains two merging nuclei, with the primary galaxy likely of early type. Given the high molecular content we find in the system ($M_{\rm{mol}}\approx9^{+19}_{-6}\times 10^8$ \msun), we speculate that the secondary galaxy was a disk galaxy which brought in the majority of the gas. 
In this scenario, the cold gas from the secondary was disrupted during the encounter.  Roughly a third of the gas sank into the primary nucleus and formed a compact disk (N nucleus, $r\approx 300$ pc, $M_{\rm{mol}}\approx3^{+6}_{-2}\times 10^8$ \msun), while half of the gas is now in an extended tidal tail (W arm, $r\approx$ 5 kpc, $M_{\rm{mol}}\approx5^{+11}_{-3}\times 10^8$ \msun), which is one of the most massive molecular tidal features known.  

In addition to a compact disk, we find a red-shifted high-velocity component deviating from rotation at the N nucleus. It has a compact size of $\sim 300$ pc and a velocity of $\sim 500$ \kms. 
Although the origin of this gas is not clear, we suspect it is an outflow driven by the AGN. Our estimated star formation rate limits, SFR $< 16$ \msun~yr$^{-1}$ from the CO molecular content and SFR $< 21$ \msun~yr$^{-1}$ from the FIR SED decomposition, disfavor star formation as the driving mechanism. 
A molecular outflow with $\dot{M}\approx350$ \msun~yr$^{-1}$ could expel all gas in the N nucleus in $\sim$ 1 Myr and all the molecular gas in the system in $\sim$ 3 Myr. 
It is one of the most compact CO molecular outflows discovered, and the corresponding dynamical time is short $\sim 0.6$ Myr.  
The kinetic luminosity of this outflow $\dot{E}\approx3\times10^{43}$~\ergs~is only $\sim 0.3\%$ of the AGN bolometric luminosity $\lbol\approx10^{46}$ \ergs. 
Its low momentum boost rate $\dot{p}/(\lbol/c)\approx$ 3 is consistent with the prediction that compact AGN outflows tend to be momentum conserving \citep{2012ApJ...745L..34Z}. 
The compact molecular outflow discovered here and the extended ionized outflow \citep{2012ApJ...746...86G} are likely induced from two episodes of AGN activity, varying on a time scale of $10$ Myr.

SDSS J1356+1026 is an example of a molecular outflow that can effectively disturb or deplete the molecular reservoir in the galaxy. 
Molecular outflows identified by high velocity OH absorption features are  found to be common among AGN \citep{2011ApJ...733L..16S, 2013ApJ...776...27V}.  
As demonstrated in this work, spatially resolved CO observations provide a complementary and model independent technique to measure the size and mass, and to infer the mass-lost rate of the outflow ($\dot{M}$). 
In the end, whether AGN feedback is a successful model to account for the absence of luminous blue galaxies and the cut-off of the galaxy luminosity function depends on both the frequency and the mass-lost rate of molecular outflows as a function of AGN luminosity. 
Furthermore, both molecular and ionized outflows are found to be ubiquitous in luminous quasars \citep{2013MNRAS.430.2327L,2013MNRAS.436.2576L, _2014arXiv1403.3086H}, and can coexist \citep{Davis:2012eo,2013ApJ...768...75R}, but we do not fully understand how they are related.
Also, the connection between the compactness and the low momentum boost rate of molecular outflows, as hinted by SDSS J1356+1026, requires confirmation from a larger spatially resolved sample. 
These outstanding questions could be answered with systematic and spatially resolved observations of molecular gas in a sample of AGN covering a range of luminosities and ionized outflow properties. 
 As demonstrated in this work, with its great sensitivity and resolution, ALMA is well suited for resolving molecular outflows on sub-kpc scales to as far as $z=0.1$,  and is therefore a powerful tool to bring our understanding of AGN feedback to the next level.

{\bf Acknowledgements:}
We thank the referee for the useful comments and Andreea Petric and Luis Ho for kindly providing the \emph{Herschel} photometry. A. Sun is thankful for Brandon Hensley, Jerry Ostriker, Jim Gunn, Paul Ho, Hauyu Baobab Liu, Satoki Matsushita, Sylvain Veilleux, Timothy Davis, and Joan Wrobel for constructive discussions. A. Sun was partially supported by the NRAO Student Observing Support grant SOSPA0-013.

% ALMA
%https://almascience.nrao.edu/alma-data
This paper makes use of the following ALMA data: ADS/JAO.ALMA\#2011.0.00652.S and 2012.1.00797.S. ALMA is a partnership of ESO (representing its member states), NSF (USA) and NINS (Japan), together with NRC (Canada) and NSC and ASIAA (Taiwan), in cooperation with the Republic of Chile. The Joint ALMA Observatory is operated by ESO, AUI/NRAO and NAOJ.
The National Radio Astronomy Observatory is a facility of the National Science Foundation operated under cooperative agreement by Associated Universities, Inc.
% WISE
% http://wise.ssl.berkeley.edu/astronomers.html
This publication makes use of data products from the Wide-field Infrared Survey Explorer, which is a joint project of the University of California, Los Angeles, and the Jet Propulsion Laboratory/California Institute of Technology, funded by the National Aeronautics and Space Administration.
% Herschel 
% http://herschel.esac.esa.int/PublishingRulesGuidelines.shtml
\emph{Herschel} is an ESA space observatory with science instruments provided by European-led Principal Investigator consortia and with important participation from NASA.

%%%%%%%%%%%%%%%%%%%%% Bibliography %%%%%%%%%%%%%%%%%%%%

\end{document}